\documentclass[proceedings]{JHEP3}
\usepackage{eurosym}
\usepackage{amsfonts}
\usepackage{amsmath}
\usepackage{epsfig}

\newbox\mybox

\newcommand\fverb{\setbox\mybox=\hbox\bgroup\verb}
\newcommand\fverbdo{\egroup\medskip\noindent\fbox{\unhbox\mybox}\ }
\newcommand\fverbit{\egroup\item[\fbox{\unhbox\mybox}]}
\conference{Perturbative approach for strong and weakly coupled time-dependent for NHQS}
\abstract{We propose a perturbative approach to determine the time-dependent Dyson map and the metric operator associated with time-dependent 
non-Hermitian Hamiltonians. We apply the method to a pair of explicitly time-dependent two dimensional harmonic oscillators that are weakly coupled to each other in a PT-symmetric fashion
and to the strongly coupled explicitly time-dependent negative quartic anharmonic oscillator potential. We demonstrate that once the perturbative Ansatz is set up the coupled differential equations resulting 
order by order may be solved recursively in a constructive manner, thus bypassing the need for making any guess for the Dyson map or the metric operator. Exploring the 
ambiguities in the solutions of the order by order differential equations naturally leads to a whole set of inequivalent solutions 
for the Dyson maps and metric operators implying different physical behaviour as demonstrated for the expectation values of the time-dependent energy operator.}

\title{Perturbative approach for strong and weakly coupled time-dependent
non-Hermitian quantum systems}
\author{Andreas Fring and Rebecca Tenney \\
Department of Mathematics, City University London,\\
Northampton Square, London EC1V 0HB, UK\\
E-mail: a.fring@city.ac.uk, rebecca.tenney@city.ac.uk}

\input{tcilatex}
\begin{document}

\section{Introduction}

The key ingredient for a physical interpretation of $\mathcal{PT}$%
-symmetric/pseudo Hermitian Hamiltonian systems requires a well defined
positive definite metric operator $\rho $. Only when this operator is
explicitly known one is in a position to define a positive definite inner
product, calculate observables together with their expectation values and
thus root the non-Hermitian theory in a well defined Hilbert space \cite%
{Bender1998,Bender2007,Mostafazadeh2010,PTbook}. In the absence of an
explicit time-dependence in the non-Hermitian Hamiltonian $H\neq H^{\dagger
} $ the metric operator $\rho $ can be determined from the time-independent
quasi-Hermiticity relation $H^{\dagger }\rho =\rho H$; in principle that is.
The metric operator can be factorised as $\rho =\eta ^{\dagger }\eta $,
where $\eta $ is often referred to as the Dyson map. The adjoint action of
this operator maps the non-Hermitian Hamiltonian to a Hermitian counterpart $%
h=h^{\dagger }$ by mean of the time-independent Dyson equation $\eta H\eta
^{-1}=h$.

For many known models the metric, and therefore the Dyson map, have been
constructed in an explicitly analytically closed form, see for instance \cite%
{Jones2006,Assis2008,Assis2009,Mostafazadeh2008,MGH}. However, in general
these \textquotedblleft solvable models\textquotedblright\ remain an
exception and one often needs to employ a perturbative approach in order to
gain some insight into the theory. Even for the classic example of a
non-Hermitian system with a real eigenvalue spectrum, complex cubic
oscillator potential $V=ix^{3}$, the metric operator is only known in a
perturbative form \cite{Bender2002,Siegl2012}. This approach has turned out
to be very successful and there are even examples for which an initially
perturbative approach has led to an exact solution with the perturbation
series terminating at a certain order, see e.g. \cite{Jones2006} for the
unstable quartic anharmonic oscillator potential $V=-x^{4}$.

When an explicit time-dependence is introduced into the Hamiltonians $%
h(t)=h(t)^{\dagger }$ and $H(t)\neq H(t)^{\dagger }$, one needs to solve the
two time-dependent Schr\"{o}dinger equations $i\hbar \partial _{t}\phi
(t)=h(t)\phi (t)$ and $i\hbar \partial _{t}\psi (t)=H(t)\psi (t)$. Assuming
that the two associated wave functions are related as $\phi (t)=\eta (t)\psi
(t)$, one easily derives \cite%
{FigueiraDeMorissonFaria2006,Mostafazadeh2007,Znojil2008,BilaAd,time7,Fring2016,Maamache2017}
that the corresponding time-dependent Dyson equation (TDDE) and
time-dependent quasi-Hermiticity relation (TDQH) acquire the forms 
\begin{equation}
h(t)=\eta (t)H(t)\eta ^{-1}(t)+i\hbar \partial _{t}\eta (t)\eta ^{-1}(t)%
\text{, \ \ \ }H^{\dagger }(t)=\rho (t)H(t)\rho ^{-1}(t)+i\hbar \partial
_{t}\rho (t)\rho ^{-1}(t),  \label{TDDE}
\end{equation}%
respectively. The novelty in the conceptual interpretation of these
equations is the fact that the non-Hermitian Hamiltonian $H(t)$, defined as
the operator that satisfies the time-dependent Schr\"{o}dinger equations,
ceases to be an observable corresponding to the energy as it is no longer
pseudo Hermitian, i.e. related to a Hermitian operator by means of a
similarity transformation. Instead, the time-dependent observable energy
operator was identified as 
\begin{equation}
\tilde{H}(t):=\eta ^{-1}(t)h(t)\eta (t)=H(t)+i\hbar \eta ^{-1}(t)\partial
_{t}\eta (t).  \label{Henergy}
\end{equation}

Evidently, to solve the two equations (\ref{TDDE}) for $\eta (t)$ and $\rho
(t)$ is more complicated than solving those for the time-independent case,
due to the presence of the additional time derivative terms. Nonetheless,
for several concrete examples exact solutions to these equations have been
constructed \cite%
{Fring2016,Fring2016a,Fring2017,Fring2017a,Fring2018a,zhang2019time,BeckyAnd2}%
. An alternative new approach, that utilizes the Lewis Riesenfeld method of
invariants \cite{Lewis1969}, has recently been developed \cite%
{Fring2018b,Fring2019b}. The advantage of this approach is that once the
invariants are constructed it becomes much simpler to solve for the
time-dependent Dyson map as there is no additional time derivative term in
the relevant equations. All these approaches rely on certain inspired
guesses for a suitable Ansatz of the metric or the Dyson map. In contrast,
the powerful feature of the time-independent perturbative approach mentioned
above is that it is entirely constructive and may be solved order by order.
So far no such perturbative approach has been developed or applied in the
time-dependent scenario. The main purpose of this paper is to develop such
an approach and explore its viability to find solutions to the equations (%
\ref{TDDE}) for $\eta (t)$ and $\rho (t)$. In particular, we seek to answer
the question of whether it is possible to apply such an approach recursively
order by order in a constructive fashion.

Besides the proposed technical advance we expect any new solution to reveal
or confirm some newly observed physical phenomena. In \cite{Fring2017b} the
remarkable and unexpected feature was found that the region in parameter
space, usually referred to as the spontaneously $\mathcal{PT}$-broken
regime, becomes physical when transgressing from the time-independent to the
time-dependent scenario. This regime is characterised by a $\mathcal{PT}$%
-symmetric Hamitonian for which the corresponding wavefunctions are $%
\mathcal{PT}$-symmetrically broken. As a consequence the energy eigenvalues
occur in complex conjugate pairs in the time-independent case. However, in
the time-dependent case the expectation values for the energy operator $%
\tilde{H}(t)$ have been found to be real for some models in that regime and
the two regimes are distinguished by qualitatively quite different types of
behaviour. Besides the energy also other physical quantities display unusual
physical behaviour, such as for instance the entropy \cite%
{Fring2019,frith2020exotic,cen2020anti}. So far all explicit solutions
constructed thereafter have confirmed these charcteristics, but up to now a
generic argument that explains the occurrence of them is still missing. We
expect that even solutions to the metric operator that are only known
perturbatively to some finite order will provide insight into these features.

Our manuscript is organized as follows: In order to set the scene and to
establish our notations we briefly recall in section 2 the perturbative
approach to determine the metric operator for time-independent non-Hermitian
Hamiltonian quantum systems. We then present our proposal for a perturbation
theory for the explicitly time-dependent scenario. In section 3 we apply the
proposed method to a pair of explicitly time-dependent two dimensional
harmonic oscillators that are weakly coupled to each other in a PT-symmetric
fashion and in section 4 to the strongly coupled negative quartic anharmonic
oscillator potential with an explicit time-dependence. In section 5 we
present our conclusions and outlook.

\section{Perturbative expansions for the metric and the Dyson map}

\subsection{Time-independent perturbation theory}

We start by recalling the time-independent perturbation theory for
determining the time-independent metric and Dyson map \cite%
{Bender2004b,mostafazadeh2005symmetric,Jones2006,FigueiraDeMorissonFaria2006}%
. We start by separating the non-Hermitian Hamiltonian into its real and
imaginary part as 
\begin{equation}
H=h_{0}+i\epsilon h_{1},\qquad \ \ \text{with}~h_{0}^{\dagger
}=h_{0},h_{1}^{\dagger }=h_{1},  \label{H}
\end{equation}%
where a real parameter $\epsilon \ll 1$ has been extracted from the
imaginary part. Assuming here for simplicity that the Dyson map is Hermitian
and of the form $\eta =e^{q/2}$, the metric operator just becomes $\rho
=\eta ^{\dagger }\eta =\eta ^{2}=e^{q}$. Making use of the standard
Baker-Campbell-Hausdorff formula 
\begin{equation}
e^{A}Be^{-A}=B+[A,B]+\frac{1}{2!}[A,[A,B]]+\frac{1}{3!}[A,[A,[A,B]]]+...
\label{eq:BCH}
\end{equation}%
and assuming that $\rho $ is invertible one can then write the
quasi-Hermiticity relation as 
\begin{equation}
H^{\dagger }=\eta ^{2}H\eta ^{-2}=H+[q,H]+\frac{1}{2!}[q,[q,H]]+\frac{1}{3!}%
[q,[q,[q,H]]]+...  \label{eq:sim}
\end{equation}%
Using the decomposition (\ref{H}) for the non-Hermitian Hamiltonian $H$ this
becomes

\begin{equation}
i[q,h_{0}]+\frac{i}{2}[q,[q,h_{0}]]+\frac{i}{3!}[q,[q,[q,h_{0}]]]+...=%
\epsilon \left( 2h_{1}+[q,h_{1}]+\frac{1}{2}[q,[q,h_{1}]]+...\right) .
\label{eq:sim2}
\end{equation}%
Expanding $q$ further as a power series in $\epsilon $ in the form 
\begin{equation}
q=\sum_{n=1}^{\infty }\epsilon ^{n}q_{n},  \label{powerexp}
\end{equation}%
one can read off the coefficients of $\epsilon ^{n}$ order by order upon
substituting (\ref{powerexp}) into (\ref{eq:sim2}). One finds that $%
[h_{0},q_{2}]=0$, so that with the choice $q_{2}=0$ all even powers in (\ref%
{powerexp}) vanish. The first three nonvanishing equations are 
\begin{align}
\lbrack h_{0},q_{1}]& =2ih_{1},  \label{fi} \\
\lbrack h_{0},q_{3}]& =\frac{i}{6}[q_{1},[q_{1},h_{1}]], \\
\lbrack h_{0},q_{5}]& =\frac{i}{6}\left(
[q_{1},[q_{3},h_{1}]]+[q_{3},[q_{1},h_{1}]]-\frac{1}{60}%
[q_{1},[q_{1},[q_{1},[q_{1},h_{1}]]]]\right) .
\end{align}%
Crucially, these equations provide a constructive scheme and can be solved
recursively order by order for $q_{1}$, $q_{2}$, \ldots\ At each order one
may add a term to $q_{i}$ that commutes with $h_{0},$ which, however, does
not change the resulting Hermitian Hamiltonian $h$. One may even find a
closed formula for the expression of $h$ involving Euler's number \cite%
{FigueiraDeMorissonFaria2006}. The metric operator is well-known not to be
unique. This feature is inherited in the time-dependent setting as will be
demonstrated below.

\subsection{Time-dependent perturbation theory}

We shall now propose a similar procedure as in the time-independent case,
however, we solve the time-dependent quasi-Hermiticity relation in (\ref%
{TDDE}) for $\rho (t)$ rather than the time-dependent Dyson equation for $%
\eta (t)$. We separate the Hamiltonian as 
\begin{equation}
H(t)=h_{0}(t)+i\epsilon h_{1}(t),\qquad \ \ \text{with}~h_{0}(t)=h_{0}^{%
\dagger }(t),h_{1}(t)=h_{1}^{\dagger }(t),
\end{equation}%
with $\epsilon \ll 1$ being a time-independent expansion parameter. The
time-dependent Dyson map is assumed to be of the form 
\begin{equation}
\eta (t)=e^{q(t)/2}=\prod_{i=1}^{j}\exp \left( \sum_{l=1}^{\infty }\epsilon
^{l}\gamma _{i}^{(l)}(t)q_{i}^{(l)}\right) .  \label{an}
\end{equation}%
At this point $\gamma _{i}^{(l)}(t)$ are time-dependent real functions, $%
q_{i}^{(l)}$ are operators and the limit $j\in \mathbb{N}$ is subject to a
suitable choice. The product in (\ref{an}) is understood to be ordered $%
\prod_{i=1}^{j}a_{i}=a_{1}a_{2}\ldots a_{j}$. For the special choice $%
q_{i}^{(1)}=q_{i}^{(2)}=...=q_{i}^{(l)}=q_{i}=q_{i}^{\dagger }$, the
exponential of the sum becomes a product of exponentials and the metric
acquires the form 
\begin{equation}
\rho (t)=\eta (t)^{\dagger }\eta (t)=\prod_{i=j}^{1}\left[
\prod_{l=k}^{1}\exp \left( \epsilon ^{l}\gamma _{i}^{(l)}q_{i}\right) \right]
\prod_{i=1}^{j}\left[ \prod_{l=1}^{k}\exp \left( \epsilon ^{l}\gamma
_{i}^{(l)}q_{i}\right) \right] ,
\end{equation}%
where $\prod_{i=j}^{1}$ denotes the reverse ordered product, that is $%
\prod_{i=j}^{1}a_{i}=a_{j}a_{j-1}\ldots a_{1}$. We have also terminated the
infinite sum in (\ref{an}) at a finite value $k$. For $k=1$ the relevant
terms in the metric are therefore identified to be 
\begin{equation}
\rho ^{(1)}(t)=\left[ \prod_{i=j}^{1}\exp \left( \epsilon \gamma
_{i}^{(1)}q_{i}\right) \right] \left[ \prod_{i=1}^{j}\exp \left( \epsilon
\gamma _{i}^{(1)}q_{i}\right) \right] .  \label{anrho}
\end{equation}%
Upon substituting this expression into the time-dependent quasi-Hermiticity
relation in (\ref{TDDE}), and expanding up to first order in $\epsilon $ we
obtain the first order differential equation 
\begin{equation}
ih_{1}+\sum_{i=1}^{j}\left( \gamma _{i}^{(1)}\left[ q_{i},h_{0}\right] +i%
\dot{\gamma}_{i}^{(1)}q_{i}\right) =0.  \label{eq:1st}
\end{equation}%
We observe from this equation that we can multiply the Dyson map by a factor
involving a time-independent phase that commutes with the Hermitian part of
the Hamiltonian. This is analogous to time-independent first order equation (%
\ref{fi}), which can be retrieved from (\ref{eq:1st}) by setting the
time-derivative terms to zero with $j=1$ and $\gamma _{1}^{(1)}=1/2$.

To second order the relevant metric results to

\begin{equation}
\rho ^{(2)}(t)=\prod_{i=j}^{1}\left[ \prod_{l=2}^{1}\exp (\epsilon
^{l}\gamma _{i}^{(l)}q_{i})\right] \prod_{i=1}^{j}\left[ \prod_{l=1}^{2}\exp
(\epsilon ^{l}\gamma _{i}^{(l)}q_{i})\right] ,  \label{2nd}
\end{equation}%
where this time we have only kept terms up to order $\epsilon ^{2}$ in the
argument of the exponential function. We substitute this into the
time-dependent quasi-Hermiticity relation in (\ref{TDDE}), and only keep
terms that are proportional to $\epsilon ^{2}$, obtaining 
\begin{align}
& 2\sum_{i=1}^{j}\left( \gamma _{i}^{(2)}[q_{i},h_{0}]+i\gamma
_{i}^{(1)}[q_{i}^{1},h_{1}]+\frac{1}{2!}(\gamma
_{i}^{(1)})^{2}[q_{i},[q_{i},h_{0}]]+i\dot{\gamma}_{i}^{(2)}q_{i}\right) 
\notag \\
& +\sum_{i=1}^{j}\left( 2\sum_{r=1,\neq i}^{j}\left( \gamma _{i}^{(1)}\gamma
_{r}^{(1)}[q_{r},[q_{i},h_{0}]]+i\dot{\gamma}_{i}^{(1)}\gamma
_{r}^{(1)}[q_{r},q_{i}]\right) +(\gamma
_{i}^{(1)})^{2}[q_{i},[q_{i},h_{0}]]\right) =0.
\end{align}%
The equations resulting from higher order in $\epsilon $ can be derived in a
similar fashion. Similar to the time-independent case, these equations can
be solved recursively order by order. In contrast, we find here that the
even ordered equations are also important, as will be demonstrated below.

Some remarks are in order with regards to the Ansatz made for the
perturbative series. First of all we assumed here that $\eta (t)$ is
Hermitian, which is not necessary and in fact implies that we are missing
some of the solutions as we shall see below. The second point to notice is
that we have not made any assumptions about the operators in the
exponentials, which are in turn determined by (\ref{eq:1st}), (\ref{2nd})
and the corresponding higher order equations. Nonetheless, we made some
assumptions about the form of the products. The factorized form is motivated
by the fact that we need to compute time derivatives of these expansions,
which makes expressions of the form $e^{A(t)+B(t)}$ with non vanishing
commutators $[A(t),B(t)]$ unsuitable. We also need to make an assumption
about the limits in the product. Let us now demonstrate for a concrete
example that the recursive solutions of the order by order equations (\ref%
{eq:1st}), (\ref{2nd}), \ldots ~do indeed lead to meaningful solutions of
the time-dependent quasi-Hermiticity relation in (\ref{TDDE}).

\section{Time-dependent coupled non-Hermitian harmonic oscillators}

As a starting point to demonstrate the effectiveness of this perturbative
approach we shall consider the following pair of time-dependent harmonic
oscillators with a Hermitian and a non-Hermitian coupling term 
\begin{equation}
H(t)=\frac{a(t)}{2}(p_{x}^{2}+x^{2})+\frac{b(t)}{2}(p_{y}^{2}+y^{2})+i\frac{%
\lambda (t)}{2}(xy+p_{x}p_{y})+\frac{\mu (t)}{2}(xp_{y}-yp_{x}),
\end{equation}%
involving the time-dependent coefficient functions $a(t)$, $b(t)$, $\lambda
(t)$, $\mu (t)\in \mathbb{R}$. This non-Hermitian Hamiltonian is symmetric
with respect to two different $\mathcal{PT}$-transformations, $[\mathcal{PT}%
_{\pm },H]=0$, where the antilinear maps are given by, $\mathcal{PT}_{\pm
}:x\rightarrow \pm x,y\rightarrow \mp y,p_{x}\rightarrow \mp
p_{x},p_{y}\rightarrow \pm p_{y},i\rightarrow -i$. It generalizes a system
previously studied in \cite{Fring2018b} for $\mu =0$, $a=b$ and can be
re-expressed in terms of Hermitian generators, $K_{i}^{\dagger }=K_{i}$, 
\begin{equation}
K_{1}=\frac{1}{2}(p_{x}^{2}+x^{2}),\quad K_{2}=\frac{1}{2}%
(p_{y}^{2}+y^{2}),\quad K_{3}=\frac{1}{2}(xy+p_{x}p_{y}),\quad K_{4}=\frac{1%
}{2}(xp_{y}-yp_{x}),
\end{equation}%
forming a closed algebra with commutation relations 
\begin{align}
& \left[ K_{1},K_{2}\right] =0, & \left[ K_{1},K_{3}\right] =iK_{4},\qquad & %
\left[ K_{1},K_{4}\right] =-iK_{3},  \notag \\
& \left[ K_{2},K_{3}\right] =-iK_{4},~~~ & \left[ K_{2},K_{4}\right]
=iK_{3},\qquad & \left[ K_{3},K_{4}\right] =i(K_{1}-K_{2})/2.  \label{alg}
\end{align}%
Thus we may rewrite the Hamiltonian $H(t)$ in terms of these generators
simply as 
\begin{equation}
H(t)=a(t)K_{1}+b(t)K_{2}+i\lambda (t)K_{3}+\mu (t)K_{4}.  \label{eq:masterh}
\end{equation}%
Denoting $c(t):=a(t)-b(t)$, we shall be considering the three different
cases for $H(t)$, characterized as: 
\begin{align}
& \text{case 1}:\quad c(t)=0\quad \text{and}\quad \mu (t)=0, \\
& \text{case 2}:\quad c(t)\neq 0\quad \text{and}\quad \mu (t)=0, \\
& \text{case 3}:\quad c(t)=0\quad \text{and}\quad \mu (t)\neq 0.
\end{align}%
The first order perturbation equation (\ref{eq:1st}) that needs to be
satisfied has many different types of solutions for each of these cases.
Therefore we shall present the different solutions in separate sections
below. We will also discuss the possibility of $\eta ^{\dagger }\neq \eta $
captured by letting some of the coefficient functions $\gamma _{i}^{(l)}$ to
be purely imaginary.

As noticed in \cite{Fring2017, Fring2018b}, an interesting feature of the
explicitly time-dependent systems is that the spontaneously broken regime of
the time-independent system becomes physical. To see whether this is also
the case here we briefly discuss the time-independent version of the
Hamiltonian (\ref{eq:masterh}) with $\dot{a}=\dot{b}=\dot{\lambda}=\dot{\mu}%
=0$ in order to create a benchmark for the $\mathcal{PT}$-broken and $%
\mathcal{PT}$-symmetric regions in the parameter space. Taking the Dyson map
to be of the form 
\begin{equation}
\eta =\exp (\theta K_{4}),\qquad \text{with \ \ }\theta =\func{arctanh}%
\left( -\frac{\lambda }{c}\right) ,
\end{equation}%
and acting adjointly on $H$ leads to the Hermitian Hamiltonian 
\begin{equation}
h=\eta H\eta ^{-1}=\frac{1}{2}\left( a+b\right) (K_{1}+K_{2})+\frac{1}{2}%
\sqrt{c^{2}-\lambda ^{2}}(K_{1}-K_{2})+\mu K_{4},  \label{eq:herm1}
\end{equation}%
with eigenvalues 
\begin{equation}
E_{n,m}=\frac{1}{2}(1+n+m)(a+b)+\frac{1}{2}(n-m)\sqrt{c^{2}-\lambda ^{2}}%
\sqrt{1+\frac{\mu ^{2}}{c^{2}-\lambda ^{2}}}.
\end{equation}%
We notice for the cases 1 and 3, that is when $c=0$, the Dyson map is
ill-defined and also the eigenvalues are complex so that these two cases are
always in the spontaneously broken $\mathcal{PT}$-regime. For case 2 we
identify a $\mathcal{PT}$-symmetric regime when $|\lambda |<|c|$ and a
spontaneously broken regime otherwise. Let us now demonstrate that the
spontaneously broken $\mathcal{PT}$-regimes can become physical when an
explicit time-dependence is introduced.

We need to treat the cases 1 and 2 separately from the case 3, as we find
that the perturbative expansions for the metric have no common overlap.

\subsection{Metric and Dyson maps with $\protect\mu (t)=0$, cases 1 and 2}

We will now show how the above perturbative equations can be solved
systematically order by order in $\epsilon $. We treat here the
non-Hermitian term as a small perturbation and set $\lambda (t)\rightarrow
\epsilon \lambda (t)$ with $\epsilon \ll 1$. When succeeding in constructing
a complete infinite series we may set $\epsilon $ back to $1$. Focusing at
first on the cases 1 and 2 with $\mu (t)=0$, the first order equation (\ref%
{eq:1st}) for the Hamiltonian $H(t)$ in (\ref{eq:masterh}) becomes 
\begin{equation}
i\lambda (t)K_{3}+\sum_{i=1}^{j}\left( \gamma _{i}^{(1)}\left[
q_{i},a(t)K_{1}+b(t)K_{2}\right] +i\dot{\gamma}_{i}^{(1)}q_{i}\right) =0.
\label{first}
\end{equation}%
When compared to the corresponding time-independent equation (\ref{fi}), we
notice that besides having to satisfy the commutative structure, the
coefficient functions are not just a set of functions of the parameters in
the model, but correspond now to a system of coupled differential equations.
Having the options in (\ref{first}) to take $j\in \{1,2,3,4\}$ with $%
q_{i}\in \{K_{1},K_{2},K_{3},K_{4}\}$, the first order equation becomes%
\begin{equation}
i\left( \lambda +c\gamma _{1}^{(1)}+\dot{\gamma}_{2}^{(1)}\right)
K_{3}+i\left( \dot{\gamma}_{1}^{(1)}-c\gamma _{2}^{(1)}\right) K_{4}+i\dot{%
\gamma}_{3}^{(1)}K_{1}+i\dot{\gamma}_{4}^{(1)}K_{2}=0.  \label{cv}
\end{equation}%
Thus setting the coefficients of all $K_{i}$ in (\ref{cv}) to zero, we
obtain two coupled first order equations for $\gamma _{1}^{(1)}$ and $\gamma
_{2}^{(1)}$. Moreover, we conclude that $\gamma _{3}^{(1)}$ and $\gamma
_{4}^{(1)}$ are time-independent. As our goal is to find a time-dependent
metric and Dyson map we set them both to zero $\gamma _{3}^{(1)}=\gamma
_{4}^{(1)}=0$. Having now fixed $j=2$ and the corresponding $q_{1}=K_{4}$, $%
q_{2}=K_{3}$, we can simply evaluate the higher order equations obtaining
the constraints by setting the coefficient functions to zero. The first
equation contains the key foundational structure for the entire series.

We proceed now in this manner to the higher order equations.

\subsubsection{Hermitian $\protect\eta $ with $q_{1}=K_{4}$ and $q_{2}=K_{3}$%
}

Keeping now the choice of the $q_{i}$ as indicated above, we derive the
differential equations to be satisfied at each order in $\epsilon $. The
first five orders of the equations to be satisfied for the $\gamma
_{1}^{(l)}(t)$ are

\begin{align}
& \epsilon ^{1}:\quad \dot{\gamma}_{1}^{(1)}=c\gamma _{2}^{(1)},  \label{h1}
\\
& \epsilon ^{2}:\quad \dot{\gamma}_{1}^{(2)}=c\gamma _{2}^{(2)}, \\
& \epsilon ^{3}:\quad \dot{\gamma}_{1}^{(3)}=c\left[ \frac{1}{6}\left(
\gamma _{2}^{(1)}\right) ^{3}+\gamma _{2}^{(3)}\right] , \\
& \epsilon ^{4}:\quad \dot{\gamma}_{1}^{(4)}=c\left[ \frac{1}{2}\left(
\gamma _{2}^{(1)}\right) ^{2}\gamma _{2}^{(2)}+\gamma _{2}^{(4)}\right] , \\
& \epsilon ^{5}:\quad \dot{\gamma}_{1}^{(5)}=c\left[ \frac{1}{120}\left(
\gamma _{2}^{(1)}\right) ^{5}+\frac{1}{2}\gamma _{2}^{(1)}\left( \gamma
_{2}^{(2)}\right) ^{2}+\frac{1}{2}\left( \gamma _{2}^{(1)}\right) ^{2}\gamma
_{2}^{(3)}+\gamma _{2}^{(5)}\right] .  \label{h5}
\end{align}

\noindent For $\gamma _{2}(t)$ we obtain the first order differential
equations 
\begin{align}
& \epsilon ^{1}:\quad \dot{\gamma}_{2}^{(1)}=-c\gamma _{1}^{(1)}-\lambda ,
\label{g2} \\
& \epsilon ^{2}:\quad \dot{\gamma}_{2}^{(2)}=-c\gamma _{1}^{(2)}, \\
& \epsilon ^{3}:\quad \dot{\gamma}_{2}^{(3)}=c\left[ \frac{1}{3}\left(
\gamma _{1}^{(1)}\right) ^{3}-\gamma _{1}^{(3)}-\frac{1}{2}\gamma
_{1}^{(1)}\left( \gamma _{2}^{(1)}\right) ^{2}\right] , \\
& \epsilon ^{4}:\quad \dot{\gamma}_{2}^{(4)}=c\left[ \left( \gamma
_{1}^{(1)}\right) ^{2}\gamma _{1}^{(2)}-\gamma _{1}^{(4)}-\frac{1}{2}\gamma
_{1}^{(2)}\left( \gamma _{2}^{(1)}\right) ^{2}-\gamma _{1}^{(1)}\gamma
_{2}^{(1)}\gamma _{2}^{(2)}\right] , \\
& \epsilon ^{5}:\quad \dot{\gamma}_{2}^{(5)}=c\left[ \gamma _{1}^{(1)}\left(
\gamma _{1}^{(2)}\right) ^{2}-\frac{2}{15}\left( \gamma _{1}^{(1)}\right)
^{5}+\left( \gamma _{1}^{(1)}\right) ^{2}\gamma _{1}^{(3)}-\gamma _{1}^{(5)}+%
\frac{1}{6}\left( \gamma _{1}^{(1)}\right) ^{3}\left( \gamma
_{2}^{(1)}\right) ^{2}\right.  \label{g5} \\
& \qquad \qquad \quad \!\!\left. -\frac{1}{24}\gamma _{1}^{(1)}\left( \gamma
_{2}^{(1)}\right) ^{4}-\frac{1}{2}\gamma _{1}^{(3)}\left( \gamma
_{2}^{(1)}\right) ^{2}-\gamma _{1}^{(2)}\gamma _{2}^{(1)}\gamma _{2}^{(2)}-%
\frac{1}{2}\gamma _{1}^{(1)}\left( \gamma _{2}^{(2)}\right) ^{2}-\gamma
_{1}^{(1)}\gamma _{2}^{(1)}\gamma _{2}^{(3)}\right] .  \notag
\end{align}%
These equations reveal the underlying structure that distinguishes the
different cases. Whilst the equations look rather complex, they contain all
the information that can be used to obtain the solutions up to fifth order
that can even be extrapolated to the exact solutions.

\paragraph{From perturbation theory to the exact Dyson map and Hermitian
Hamiltonians}

We shall now demonstrate how to use these equations to obtain the Dyson map
and hence the metric. Proceeding similarly as for the first order equation (%
\ref{cv}), we may solve the set of equations (\ref{h1})-(\ref{h5}), (\ref{g2}%
)-(\ref{g5}) recursively order by order to obtain the explicit expressions
for the coefficient functions $\gamma _{1}^{(i)}(t)$ and $\gamma
_{2}^{(i)}(t)$, $i=1,2,\ldots $ We will not report these expressions here.
In the next step we extrapolate from the first terms by trying to identify a
combination of standard functions whose Taylor expansion matches the first
terms in the perturbative series.

\noindent \underline{For case 1}, when $c(t)=0$, we notice from (\ref{cv})
that also $\dot{\gamma}_{1}^{(1)}=0$ when requiring Hermiticity of $h$. As
the Hermitian part of the Hamiltonian $H(t)$ is given by $%
h_{0}(t)=a(t)(K_{1}+K_{2})$, we now have $[h_{0}(t),K_{i}]$ so that all of
the generators in this algebra commute with $h_{0}(t)$. As a consequence of
this we observe that all orders of the perturbation equations disappear
except for one. This is also seen by setting $c=0$ in (\ref{h1})-(\ref{g5})
so that the only relevant equation left is 
\begin{equation}
\dot{\gamma}_{2}^{(1)}(t)=-\lambda (t).
\end{equation}%
Hence, we easily obtain the exact solution 
\begin{equation*}
\gamma _{1}^{(1)}(t)=\gamma _{1}(t)=k_{1},\qquad \gamma _{2}^{(1)}(t)=\gamma
_{2}(t)=-\int^{t}\lambda (s)ds+k_{2},
\end{equation*}%
with two integration constants $k_{1}$, $k_{2}$.

\noindent \underline{For case 2}, when $c(t)\neq 0$, all of the right hand
sides of the differential equations are proportional to $c(t)$, except for
the one for $\dot{\gamma}_{2}^{(1)}(t)$ in (\ref{g2}). Assuming $\lambda (t)$
to be a real multiple of $c(t)$ the equations become fully integrable and we
are able to solve the equations order by order, even leading to an exact
solution. Keeping for instance terms up to fifth order we obtain%
\begin{equation}
\left[ \dot{\gamma}_{1}(t)\right] ^{\left[ 5\right] }=\sum_{i=1}^{5}\epsilon
^{i}\dot{\gamma}_{1}^{(i)}(t)=c(t)\left[ \epsilon \sinh \left(
\sum_{i=1}^{5}\epsilon ^{i}\gamma _{2}^{(i)}(t)\right) \right] ^{\left[ 5%
\right] }=c(t)\left\{ \epsilon \sinh \left[ \gamma _{2}(t)\right] \right\} ^{%
\left[ 5\right] },  \label{ex1}
\end{equation}%
and 
\begin{align}
\left[ \dot{\gamma}_{2}(t)\right] ^{\left[ 5\right] }&
=\sum_{i=1}^{5}\epsilon ^{i}\dot{\gamma}_{2}^{(i)}(t)=-\lambda
(t)-c(t)\left\{ \epsilon \left[ \cosh \left( \sum_{i=1}^{5}\epsilon
^{i}\gamma _{2}^{(i)}(t)\right) \right] \left[ \tanh \left(
\sum_{i=1}^{5}\epsilon ^{i}\gamma _{1}^{(i)}(t)\right) \right] \right\} ^{%
\left[ 5\right] }  \notag \\
& =-\lambda (t)-c(t)\left( \epsilon \cosh [\gamma _{2}(t)]\tanh [\gamma
_{1}(t)]\right) ^{\left[ 5\right] }.  \label{ex2}
\end{align}%
Here the superscript $\left[ 5\right] $ means we only retain terms up to
order 5 in $\epsilon $. In fact, we have verified the validity of the closed
form to eleventh order, by extending and solving the sets of equations (\ref%
{h1})-(\ref{h5}) and (\ref{g2})-(\ref{g5}).

Assuming now the expressions in (\ref{ex1}) and (\ref{ex2}) to be exact, we
may set $\epsilon =1$ and subsequently solve them for $\gamma _{1}(t)$ and $%
\gamma _{2}(t)$. Letting $\lambda (t)$ be any real multiple of $c(t)$, that
is 
\begin{equation}
c(t)=p\lambda (t)\qquad \text{where}\qquad p\in \mathbb{R},  \label{lam}
\end{equation}%
we are able to solve the relevant equations exactly and express $\gamma _{2}$
as a function of $\gamma _{1}$ as 
\begin{equation}
\gamma _{2}(\gamma _{1})=\pm \func{arccosh}\left\{ -\frac{1}{2}\func{sech}%
(\gamma _{1})\left[ k_{1}+\frac{2}{p}\sinh (\gamma _{1})\right] \right\} ,
\label{g21}
\end{equation}%
with $k_{1}$ being an integration constant. Relation (\ref{g21}) is obtained
by integrating $\dot{\gamma}_{2}/\dot{\gamma}_{1}=\partial \gamma
_{2}/\partial \gamma _{1}$ with respect to $\gamma _{1}$. Parameterizing $%
\gamma _{1}(t)$ by a new function $\chi (t)$ as%
\begin{equation}
\gamma _{1}=\func{arcsinh}\left( \chi \right) ,
\end{equation}%
the two differential equations for $\dot{\gamma}_{1}(t)$ and $\dot{\gamma}%
_{2}(t)$ can be converted into the linear second order equation entirely in $%
\chi $ 
\begin{equation}
\ddot{\chi}-\frac{\dot{\lambda}}{\lambda }~\dot{\chi}+(p^{2}-1)\lambda
^{2}\chi =k_{1}\frac{p}{2}\lambda ^{2}.  \label{EP}
\end{equation}

\noindent We solve equation (\ref{EP}) by%
\begin{equation}
\chi (t)=\frac{e^{-2\sqrt{1-p^{2}}\left( k_{2}-\frac{1}{2}\int^{t}\lambda
(s)ds\right) }}{4\left( 1-p^{2}\right) }\left[ \left( e^{2\sqrt{1-p^{2}}%
\left( k_{2}-\frac{1}{2}\int^{t}\lambda (s)ds\right) }-pk_{1}\right)
^{2}+(k_{1}^{2}-4)\left( 1-p^{2}\right) \right] .~  \label{solaux1}
\end{equation}%
Notice that in fact we are solving the two first order equations for $\dot{%
\gamma}_{1}(t)$ and $\dot{\gamma}_{2}(t)$, so that there are only two
integration constants and no additional linear independent solution for the
second order equation (\ref{EP}). We have to impose here $\left\vert
p\right\vert <1$ to ensure the reality of $\chi $ and hence $\gamma _{2}$, $%
\gamma _{1}$.

Having obtained an exact Dyson map, we can envoke the first equation in (\ref%
{TDDE}) and compute the Hermitian counterparts to $H(t)$, which consists of
two decoupled harmonic oscillators in both cases 1 and 2 
\begin{equation}
h(t)=f_{+}(t)K_{1}+f_{-}(t)K_{2}.  \label{decharm}
\end{equation}%
For case 1 we find $f_{\pm }(t)=a$ and for case 2 we obtain%
\begin{equation}
f_{\pm }(t)=b+\frac{p\lambda }{2}\mp \frac{\lambda (2\chi +pk_{1})}{4(1+\chi
^{2})}.
\end{equation}%
We may also compute real time-dependent energy expectation values from these
expressions as will be shown below.

\subsubsection{Non-Hermitian $\protect\eta $ with $q_{1}=K_{4}$ and $%
q_{2}=K_{1},K_{2}$}

Making now the choice $q_{1}=K_{4}$, $q_{2}=K_{1},K_{2}\,$\ the perturbative
expansion yields $\dot{\gamma}_{1}^{(l)}=\dot{\gamma}_{2}^{(l)}=0$, so that
the entire metric becomes time-independent. However, $\eta $ does not have
to be Hermitian as assumed in the Ansatz (\ref{an}). Thus allowing $\gamma
_{i}^{(l)}\in \mathbb{C}$ in general, we now modify the Ansatz to $\gamma
_{1}^{(\ell )}\in \mathbb{R}$, \ $\ell =1,2,\ldots $, $\gamma _{2}^{(\ell
)}\in i\mathbb{R}$, \ $\ell =0,1,2,\ldots $, $\gamma _{3}^{(\ell )}=\gamma
_{4}^{(\ell )}=0.$ The perturbative constraints up to order $\epsilon ^{3}$
then read%
\begin{align}
& \epsilon ^{1}:\quad \dot{\gamma}_{1}^{(1)}=\lambda \sin \left( \gamma
_{2}^{(0)}\right) , \\
& \epsilon ^{2}:\quad \dot{\gamma}_{1}^{(2)}=\lambda \gamma _{2}^{(1)}\cos
\left( \gamma _{2}^{(0)}\right) , \\
& \epsilon ^{3}:\quad \dot{\gamma}_{1}^{(3)}=\lambda \gamma _{2}^{(2)}\cos
\left( \gamma _{2}^{(0)}\right) -\frac{1}{2}\lambda \left( \gamma
_{2}^{(1)}\right) ^{2}\sin \left( \gamma _{2}^{(0)}\right) ,
\end{align}

\noindent and for $\gamma _{2}(t)$ we obtain 
\begin{align}
& \epsilon ^{1}:\quad \dot{\gamma}_{2}^{(0)}=c+\lambda \frac{\cos \left(
\gamma _{2}^{(0)}\right) }{\gamma _{1}^{(1)}}, \\
& \epsilon ^{2}:\quad \dot{\gamma}_{2}^{(1)}=-\frac{\lambda }{\gamma
_{1}^{(1)}}\left[ \frac{\gamma _{1}^{(2)}}{\gamma _{1}^{(1)}}\cos \left(
\gamma _{2}^{(0)}\right) +\gamma _{2}^{(1)}\sin \left( \gamma
_{2}^{(0)}\right) \right] , \\
& \epsilon ^{3}:\quad \dot{\gamma}_{2}^{(2)}=\frac{\lambda }{\gamma
_{1}^{(1)}}\left\{ \left[ \frac{\left( \gamma _{1}^{(1)}\right) ^{2}}{3}%
+\left( \frac{\gamma _{1}^{(2)}}{\gamma _{1}^{(1)}}\right) ^{2}-\frac{\gamma
_{1}^{(3)}}{\gamma _{1}^{(1)}}-\frac{\gamma _{2}^{(1)}}{2}\right] \cos
\left( \gamma _{2}^{(0)}\right) \right. \\
\qquad \qquad & \qquad \qquad \qquad +\left. \left[ \frac{\gamma
_{1}^{(2)}\gamma _{2}^{(1)}}{\gamma _{1}^{(1)}}-\gamma _{2}^{(2)}\right]
\sin \left( \gamma _{2}^{(0)}\right) \right\} .  \notag
\end{align}

\paragraph{From perturbation theory to the exact Dyson map and Hermitian
Hamiltonians}

Once again we may solve these equations order by order for the coefficient
functions $\gamma _{i}^{(\ell )}$ and subsequently try to extrapolate the
series to all orders. We find the exact constraining equations for $\gamma
_{1}(t)$ and $\gamma _{2}(t)$ by demanding the non-Hermitian terms in $h(t)$
to vanish%
\begin{equation*}
\dot{\gamma}_{1}=\lambda \sin \left( \gamma _{2}\right) ,\quad ~~\text{%
and\quad ~}\dot{\gamma}_{2}=c+\lambda \cos \left( \gamma _{2}\right) \coth
(\gamma _{1}).
\end{equation*}%
We may now solve these equations separately in each case.

\noindent \underline{For case 1}\textbf{\ }with\textbf{\ }$q_{2}=K_{1}$, we
can solve for $\gamma _{1}$ in terms of $\gamma _{2}$ obtaining 
\begin{equation}
\gamma _{1}(\gamma _{2})=\func{arcsinh}\left[ k_{1}\sec (\gamma _{2})\right]
,
\end{equation}%
with integration constant $k_{1}$. By letting 
\begin{equation}
\gamma _{2}=\arctan (\chi ),
\end{equation}%
the equations for $\dot{\gamma}_{1}$ and $\dot{\gamma}_{2}$ are converted
into the linear second order differential equation%
\begin{equation}
\ddot{\chi}-\frac{\dot{\lambda}}{\lambda }\dot{\chi}-\lambda ^{2}\chi =0.
\label{e2}
\end{equation}%
We observe that the auxiliary equation (\ref{EP}) reduces to equation (\ref%
{e2}) in the limit $p\rightarrow 0$ which also holds for the solution (\ref%
{solaux1}). We have two constants of integration left after having carried
out the limit.

\noindent \underline{For case 2}\textbf{\ }with\textbf{\ }$q_{2}=K_{1}$, we
set $c(t)=p\lambda (t)$ as then the equations become solvable. In this case
it is more convenient to express $\gamma _{2}$ in terms of $\gamma _{1}$ 
\begin{equation}
\gamma _{2}(\gamma _{1})=\arccos \left[ -p\coth (\gamma _{1})-i\frac{1}{2}%
k_{1}\func{csch}(\gamma _{1})\right] ,
\end{equation}%
where $k_{1}$ is an integration constant that we set to $0$ to ensure the
reality of $\gamma _{2}$. Letting 
\begin{equation}
\gamma _{1}=\func{arccosh}\left( \chi \right) ,
\end{equation}%
the equations for $\dot{\gamma}_{1}$ and $\dot{\gamma}_{2}$ are converted
into the linear second order differential equation%
\begin{equation}
\ddot{\chi}-\frac{\dot{\lambda}}{\lambda }\dot{\chi}+(p^{2}-1)\lambda
^{2}\chi =0.  \label{e3}
\end{equation}%
We note that equations (\ref{e3}) is obtained from (\ref{EP}) in the limit $%
k_{1}\rightarrow 0$, which also holds for the solution (\ref{solaux1}). As
we have already chosen one of the integration constants, there is only one
left in this case, i.e. $k_{2}$.

After imposing the constraints, the remaining Hermitian part of the
Hamiltonian is of the same general form as the one reported in (\ref{decharm}%
), albeit with different forms for the coefficient functions%
\begin{equation}
f_{\pm }(t)=b-\frac{\lambda (\pm 1+\sqrt{1+(1+\chi ^{2})k_{1}^{2}})}{%
2(1+\chi ^{2})k_{1}},
\end{equation}%
in case 1 and 
\begin{equation}
f_{\pm }(t)=b+\frac{p\lambda \chi }{2(\chi \mp 1)},
\end{equation}%
in case 2, respectively.

\subsubsection{Further choices that lead to exact Dyson maps and Hermitian $%
h(t)$}

Having made a distinction in the setup of the perturbative treatment between
Hermitian and non-Hermitian Dyson maps, there are further possible choices
within these two frameworks that all lead to exactly solvable solutions. As
the procedure to find them is similar to the previous cases we present them
in a more compact form, omitting the details of the derivations. The
constraining relations arising from requiring the transformed Hamiltonian $%
h(t)$ in (\ref{TDDE}) to be Hermitian are presented in table 1. For
completeness, we also report the cases discussed already in more detail
above.

\begin{table}[h]
\begin{center}
\begin{tabular}{|c||c|c|}
\hline
$q_{1},q_{2}$ & $\dot{\gamma}_{1}(t)$ & $\dot{\gamma}_{2}(t)$ \\ \hline\hline
$K_{4},K_{3}$ & $c\sinh (\gamma _{2})$ & $-c\cosh \left( \gamma _{2}\right)
\tanh \left( \gamma _{1}\right) -\lambda $ \\ \hline
$K_{3},K_{4}$ & $~~-\lambda \cosh \left( \gamma _{2}\right) -c\sinh \left(
\gamma _{2}\right) ~~$ & $~~\left[ c\cosh \left( \gamma _{2}\right) +\lambda
\sinh \left( \gamma _{2}\right) \right] \tanh \left( \gamma _{1}\right) ~~$
\\ \hline
$K_{4},iK_{1,2}$ & $\pm \lambda \sin (\gamma _{2})$ & $\pm c\pm \lambda \cos
(\gamma _{2})\coth (\gamma _{1})$ \\ \hline
$K_{3},iK_{1,2}$ & $-\lambda \cos (\gamma _{2})$ & $\pm c+\lambda \sin
(\gamma _{2})\coth (\gamma _{1})$ \\ \hline
\end{tabular}%
\end{center}
\caption{Coupled first order differential equation constraints on the
time-dependent coefficient functions $\protect\gamma _{1}$ and $\protect%
\gamma _{2}$ in the Dyson map $\protect\eta $, for different choices of $%
q_{1}$ and $q_{2}$.}
\end{table}

\begin{table}[h]
\begin{center}
\begin{tabular}{|l||l|l|l|l|}
\hline
$q_{1},q_{2}$ & constraint & $\gamma _{1}(\chi )$ & $\gamma _{2}(\chi )$ & 
constraint \\ \hline\hline
$K_{4},K_{3}$ & $c=0$ & * & * & * \\ \hline
$K_{4},K_{3}$ & $c=p\lambda $ & $\func{arcsinh}\left( \chi \right) $ & $%
\func{arccosh}\left( -\frac{k_{1}+2p\chi }{2\sqrt{1+\chi ^{2}}}\right) $ & -$%
\frac{k_{1}+2p\chi }{2\sqrt{1+\chi ^{2}}}\leq 1$ \\ \hline
$K_{3},K_{4}$ & $c=0$ & $\func{arccosh}\left( \chi \right) $ & $\func{arcsinh%
}\left( \frac{k_{1}}{\chi }\right) $ & $\chi >1$ \\ \hline
$K_{3},K_{4}$ & $c=\lambda $ & $\func{arccosh}\left( \chi \right) $ & $\ln
\left( \frac{k_{1}}{\chi }\right) $ & $\chi >1$ \\ \hline
$K_{4},iK_{1,2}$ & $c=0$ & $\func{arcsinh}\left( k_{1}\sqrt{1+\chi ^{2}}%
\right) $ & $\pm \arctan (\chi )$ & * \\ \hline
$K_{4},iK_{1,2}$ & $c=p\lambda $ & $\func{arccosh}\left( \chi \right) $ & $%
\arccos \left( -\frac{p\chi }{\sqrt{\chi ^{2}-1}}\right) $ & $\chi >1$ \\ 
\hline
$K_{3},iK_{1,2}$ & $c=0$ & $\func{arcsinh}\left( k_{1}\sqrt{1+\chi ^{2}}%
\right) $ & $\pm \func{arccot}(\chi )$ & * \\ \hline
$K_{3},iK_{1,2}$ & $c=p\lambda $ & $\func{arccosh}\left( \chi \right) $ & $%
\arcsin \left( \frac{k_{1}\mp 2p\chi }{2\sqrt{\chi ^{2}-1}}\right) $ & $\chi
>1$ \\ \hline
\end{tabular}%
\end{center}
\caption{Parameterisation of $\protect\gamma _{1}$ and $\protect\gamma _{2}$
in terms of the auxiliary function $\protect\chi $ with additional
constraint on $c(t)$ for different choices of $q_{1}$ and $q_{2}$. The
constraints in the last column result from the parameterisation. A $\ast $
indicates no constraint.}
\end{table}

All presented solutions and cases are new, except for the Hermitian case
with $q_{1}=K_{3}$, $q_{2}=K_{4}$, $c=0$ which reproduces a solution found
in \cite{Fring2018a}, with the difference that the Dyson map we are
considering here are missing the two factors involving the time-independent $%
K_{1}$ and $K_{2}$ terms. We can proceed as above to solve the coupled
differential equations in all cases by expressing $\gamma _{1}$ as a
function of $\gamma _{2},$ or vice versa, and a subsequent integration. The
parameterization of $\gamma _{1,2}$ in terms of a new function, that we
always denote as $\chi (t)$, are not obvious and are therefore presented in
table 2. We may only solve these equations upon imposing an additional
restriction on the time-dependent functions in the Hamiltonian, which are
also reported in table 2.

We still need to determine the auxiliary function. As discussed in the
previous subsection, combining the equations for the constraints on $\gamma
_{1}$ and $\gamma _{2}$ leads to a set of second order auxiliary equations
that we present in table 3.

\begin{table}[h]
\begin{center}
\begin{tabular}{|l||l|l|}
\hline
$q_{1},q_{2}$ & constraint & auxiliary equation \\ \hline\hline
$K_{4}$, $K_{3}$ & $c=0$ & none \\ \hline
$%
\begin{array}{l}
K_{4},K_{3} \\ 
K_{3},iK_{1,2}%
\end{array}%
$ & $c=p\lambda $ & Aux$_{1}:~~\ddot{\chi}-\frac{\dot{\lambda}}{\lambda }~%
\dot{\chi}-(1-p^{2})\lambda ^{2}\chi =k_{1}\frac{p}{2}\lambda ^{2}$ \\ \hline
$K_{3,4}$, $iK_{1,2}$ & $c=0$ & Aux$_{2}:~~\ddot{\chi}-\frac{\dot{\lambda}}{%
\lambda }~\dot{\chi}-\lambda ^{2}\chi =0$ \\ \hline
$K_{4}$, $iK_{1,2}$ & $c=p\lambda $ & Aux$_{3}:~~\ddot{\chi}-\frac{\dot{%
\lambda}}{\lambda }~\dot{\chi}-(1-p^{2})\lambda ^{2}\chi =0$ \\ \hline
$K_{3}$, $K_{4}$ & $c=0$ & Aux$_{4}:~~\ddot{\chi}-\frac{\dot{\lambda}}{%
\lambda }~\dot{\chi}-\lambda ^{2}\chi =k_{1}^{2}\lambda ^{2}\frac{1}{\chi
^{3}}$ \\ \hline
$K_{3}$, $K_{4}$ & $c=\lambda $ & Aux$_{5}:~~\ddot{\chi}-\frac{\dot{\lambda}%
}{\lambda }~\dot{\chi}=k_{1}^{2}\lambda ^{2}\frac{1}{\chi ^{3}}$ \\ \hline
\end{tabular}%
\end{center}
\caption{Auxiliary equations to be satisfied by quantities in the
parameterisation of the functions $\protect\gamma _{1}$ and $\protect\gamma %
_{2}$ together with the additional constraint on $c(t)$ for different
choices of $q_{1}$ and $q_{2}$.}
\end{table}

\paragraph{Solutions to the auxiliary equations}

As the last step we disentangle the parameterisations for $\gamma _{1}$ and $%
\gamma _{2}$ by solving the auxiliary equations for $\chi $. We have
encountered one case with no restrictions at all, three types of linear
second order equations and two versions of the nonlinear Ermakov-Pinney (EP)
equation \cite{Ermakov1880,Pinney1950}

We already reported the solutions to the linear equations referred to as Aux$%
_{1}$ in table 3 in (\ref{solaux1}), from which we obtain the solution to Aux%
$_{2}$ in the limit $p\rightarrow 0$ and Aux$_{3}$ in limit $%
k_{1}\rightarrow 0$. Hence we just need to present the solutions to the
EP-equations. We find the following solutions to Aux$_{4}$ and Aux$_{5}$%
\begin{eqnarray}
\chi (t) &=&\left[ 1+(1+k_{1}^{2})\sinh ^{2}\left( k_{2}-\int^{t}\lambda
(s)ds\right) \right] ^{1/2},  \label{aux4sol} \\
\chi (t) &=&\left[ 1+\left( k_{2}-k_{1}\int^{t}\lambda (s)ds\right) ^{2}%
\right] ^{1/2},
\end{eqnarray}%
respectively.

Finally we turn to the resulting Hermitian Hamiltonian $h(t)$ that is always
of the general form of two uncoupled harmonic oscillators (\ref{decharm})
with different time-dependent coefficient functions $f_{\pm }(t)$ as
reported in table 4.

\begin{table}[h]
\begin{center}
\begin{tabular}{|l||l|l|l|}
\hline
$q_{1},q_{2}$ & constraint & $f_{\pm }(t)$ & $\eta $ \\ \hline\hline
$K_{4}$, $K_{3}$ & $c=0$ & $a$ & $\eta _{1}$ \\ \hline
$K_{4}$,$~K_{3}$ & $c=p\lambda $ & $b+\frac{p\lambda }{2}\mp \frac{\lambda
(2\chi +pk_{1})}{4(1+\chi ^{2})}$ & $\eta _{1}$ \\ \hline
$K_{3}$, $K_{4}$ & $c=0$ & $b\pm \frac{\lambda k_{1}}{2\chi ^{2}}$ & $\eta
_{2}$ \\ \hline
$K_{3}$, $K_{4}$ & $c=\lambda $ & $b+\frac{\lambda }{2}\pm \frac{\lambda
k_{1}}{2\chi ^{2}}$ & $\eta _{2}$ \\ \hline
$K_{4}$, $iK_{1}$ & $c=0$ & $b-\frac{\lambda (\pm 1+\sqrt{1+(1+\chi
^{2})k_{1}^{2}})}{2(1+\chi ^{2})k_{1}}$ & $\eta _{3}$ \\ \hline
$K_{4}$, $iK_{1}$ & $c=p\lambda $ & $b+\frac{p\lambda \chi }{2(\chi \mp 1)}$
& $\eta _{3}$ \\ \hline
$K_{4}$, $iK_{2}$ & $c=0$ & $b+\frac{\lambda (\mp 1+\sqrt{1+(1+\chi
^{2})k_{1}^{2}})}{2k_{1}(1+\chi ^{2})}$ & $\eta _{4}$ \\ \hline
$K_{4}$, $iK_{2}$ & $c=p\lambda $ & $b+p\lambda -\frac{p\lambda \chi }{%
2(\chi \pm 1)}$ & $\eta _{4}$ \\ \hline
$K_{3}$, $iK_{1}$ & $c=0$ & $b+\frac{\lambda \left[ \mp 1-\sqrt{%
1+k_{1}^{2}\left( 1+\chi ^{2}\right) }\right] }{2k_{1}\left( \chi
^{2}+1\right) }$ & $\eta _{5}$ \\ \hline
$K_{3}$, $iK_{1}$ & $c=p\lambda $ & $b+\frac{\lambda (2p\chi -k_{1})}{4(\chi
\mp 1)}$ & $\eta _{5}$ \\ \hline
$K_{3}$, $iK_{2}$ & $c=0$ & $b+\frac{\lambda \left[ \pm 1-\sqrt{%
1+k_{1}^{2}\left( 1+\chi ^{2}\right) }\right] }{2k_{1}\left( 1+\chi
^{2}\right) }$ & $\eta _{6}$ \\ \hline
$K_{3}$, $iK_{2}$ & $c=p\lambda $ & $b-\frac{\lambda (2p\chi +k_{1})}{4(\chi
\pm 1)}$ & $\eta _{6}$ \\ \hline
\end{tabular}%
\end{center}
\caption{Time-dependent coefficient in the Hermitian Hamiltonian $h(t)=$ $%
f_{+}(t)K_{1}+f_{-}(t)K_{2}$ together with the additional constraint on $%
c(t) $ for different choices of $q_{1}$ and $q_{2}$. In the last column we
report a short notation for the Dyson maps of the particular cases that we
shall use below for convenience.}
\end{table}

\subsubsection{Time-dependent eigenfunctions, energies and $\mathcal{PT}$%
-symmetry breaking}

Next we present the expectation values for the time-dependent energy
operator $\tilde{H}(t)$ as defined in equation (\ref{Henergy}). Since each
of the Hermitian Hamiltonians constructed from any of the similarity
transformations simply consists of two uncoupled harmonic oscillators (\ref%
{decharm}) with different time-dependent coefficient functions, we can
easily construct the total wavefunction as a product of the wavefunctions
for a harmonic oscillator with real time-dependent mass and frequency of the
form $\tilde{h}(t)=f(t)/2(p_{x}^{2}+x^{2})$. The latter problem was solved
originally in \cite{Pedrosa1997}. Adapting to our notation and including a
normalization constant, found in \cite{Fring2018b}, the time-dependent
wavefunction is given by 
\begin{equation}
\tilde{\phi}_{n}(x,t)=\frac{e^{i\alpha _{n}(t)}}{\sqrt{2^{n}n!\sqrt{\pi }%
\chi (t)}}\exp \left[ \left( \frac{i}{f(t)}\frac{\dot{\chi}(t)}{\chi (t)}-%
\frac{1}{\chi ^{2}(t)}\right) \frac{x^{2}}{2}\right] H_{n}\left[ \frac{x}{%
\chi (t)}\right] ,  \label{eq:wave1}
\end{equation}%
where $H_{n}\left[ x\right] $ denotes the n-th Hermite polynomial in $x$ and
the phase is given by 
\begin{equation}
\alpha _{n}(t)=-\left( n+\frac{1}{2}\right) \int_{0}^{t}\frac{f(s)}{\chi
^{2}(s)}ds.
\end{equation}%
The auxiliary function $\chi (t)$ is constrained by the dissipative
Ermakov-Pinney equation of the form 
\begin{equation}
\ddot{\chi}-\frac{\dot{f}}{f}\dot{\chi}+f^{2}\chi =\frac{f^{2}}{\chi ^{3}}.
\label{au}
\end{equation}%
Interestingly this is equation Aux$_{4}$ in table 3 with $\lambda
\rightarrow if$, $k_{1}^{2}=i$. However, the solution (\ref{aux4sol}) to Aux$%
_{4}$ reduces to 1 for these parameter choices. Instead, equation (\ref{au})
is solved by 
\begin{equation}
\chi (t)=\sqrt{\sqrt{1+c^{2}}+c\cos \left[ 2\int^{t}f(s)ds\right] },
\label{eq:ep}
\end{equation}%
with integration constant $c$. The expectation value of $K_{1}$ is given
then computed to 
\begin{equation}
\left\langle \tilde{\phi}_{n}(x,t)\right\vert K_{1}\left\vert \tilde{\phi}%
_{m}(x,t)\right\rangle =\left( n+\frac{1}{2}\right) \sqrt{1+c^{2}}\delta
_{n,m}.
\end{equation}%
Hence, the solution to the full time-dependent Schr\"{o}dinger equation for
the Hermitian Hamiltonian $h(t)$ in (\ref{decharm}) is simply the product of
the two wavefunctions in (\ref{eq:wave1}) 
\begin{equation}
\Psi _{h}^{n,m}(x,y,t)=\tilde{\phi}_{n}^{f_{+}}(x,t)\tilde{\phi}%
_{m}^{f_{-}}(y,t),
\end{equation}%
from which we calculate the instantaneous energy expectation values 
\begin{equation}
E^{n,m}(t)=\left\langle \Psi _{h}^{n,m}(t)\right\vert h(t)\left\vert \Psi
_{h}^{n,m}(t)\right\rangle =\sum\limits_{i=-,+}f_{i}(t)\left( n+\frac{1}{2}%
\right) \sqrt{1+c_{i}^{2}}.  \label{inenergy}
\end{equation}%
These expectation values are real provided $f_{\pm }(t),c_{\pm }\in \mathbb{R%
}$. For case 1 this is simply guaranteed by taking the parameter and
time-dependent functions to be real. For case 2 we can not freely choose and
have to respect the constraints resulting as a consequence of the
parameterization as reported in table 2. As the auxiliary function $\chi (t)$
must be real, the additional constraint $\left\vert p\right\vert <1$ results
from the form of the solution (\ref{solaux1}), together with $k_{1},k_{2}\in 
\mathbb{R}$.

\FIGURE{\epsfig{file=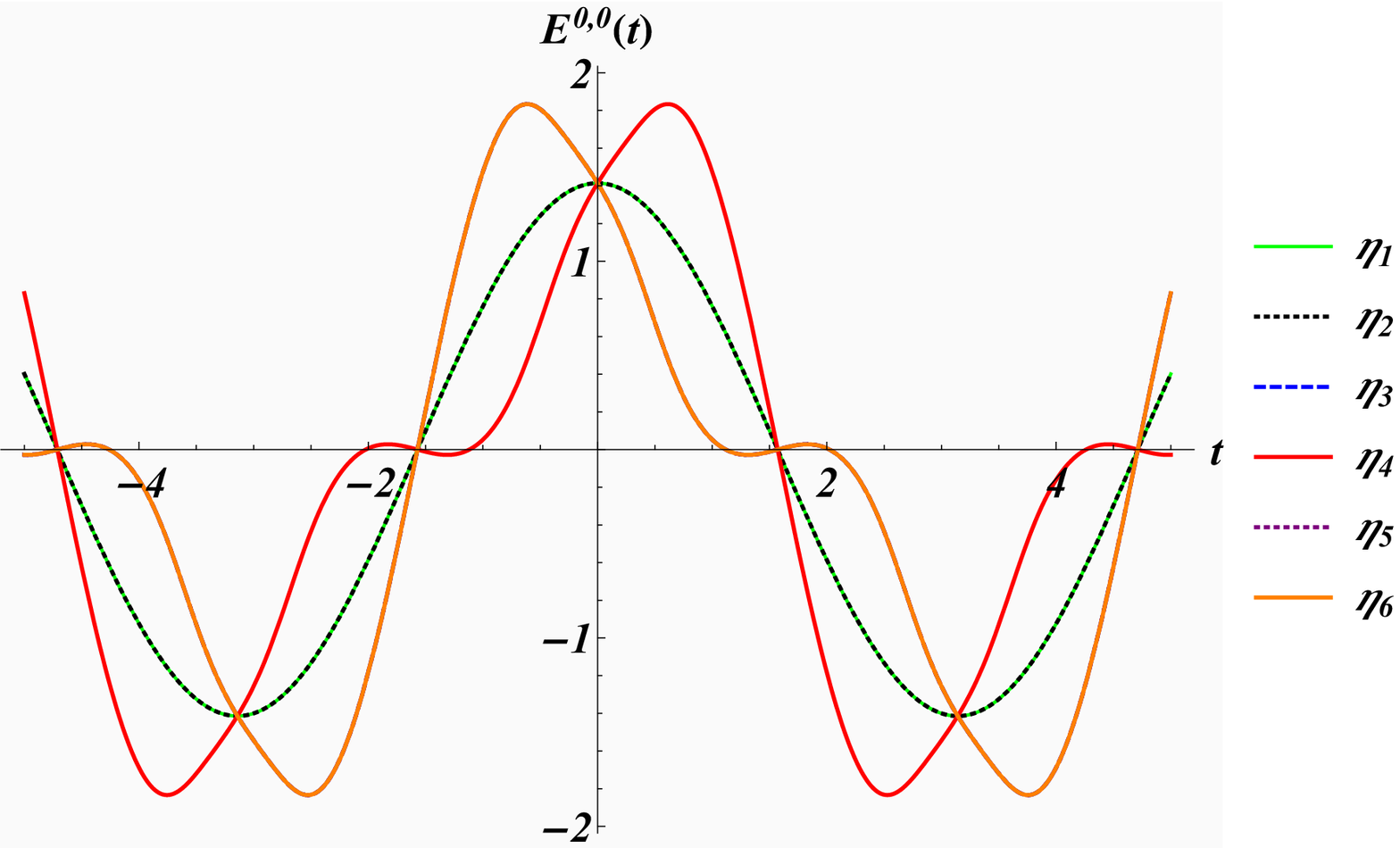, width=7.8cm} \epsfig{file=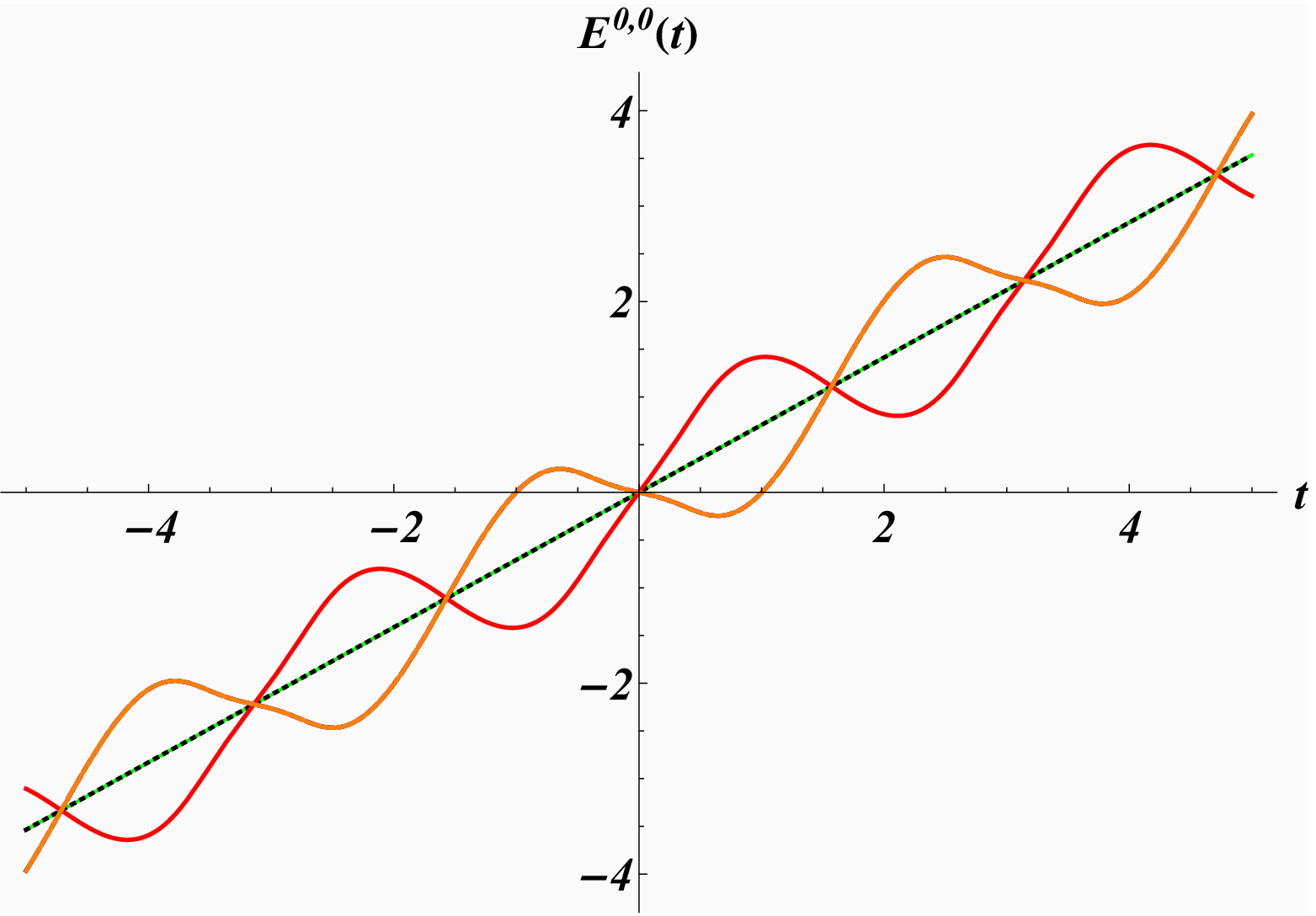, width=6.8cm}
\epsfig{file=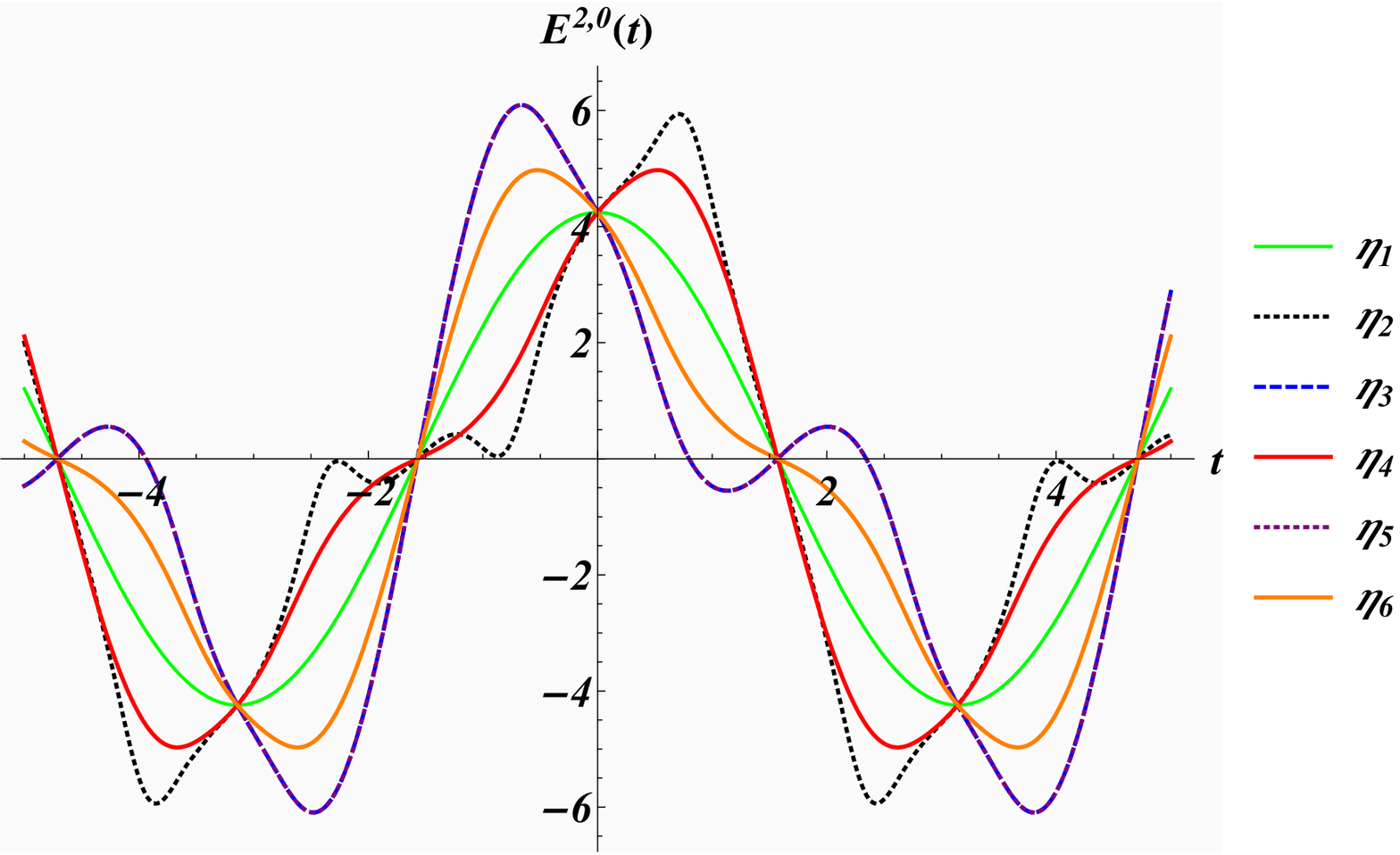, width=7.8cm} \epsfig{file=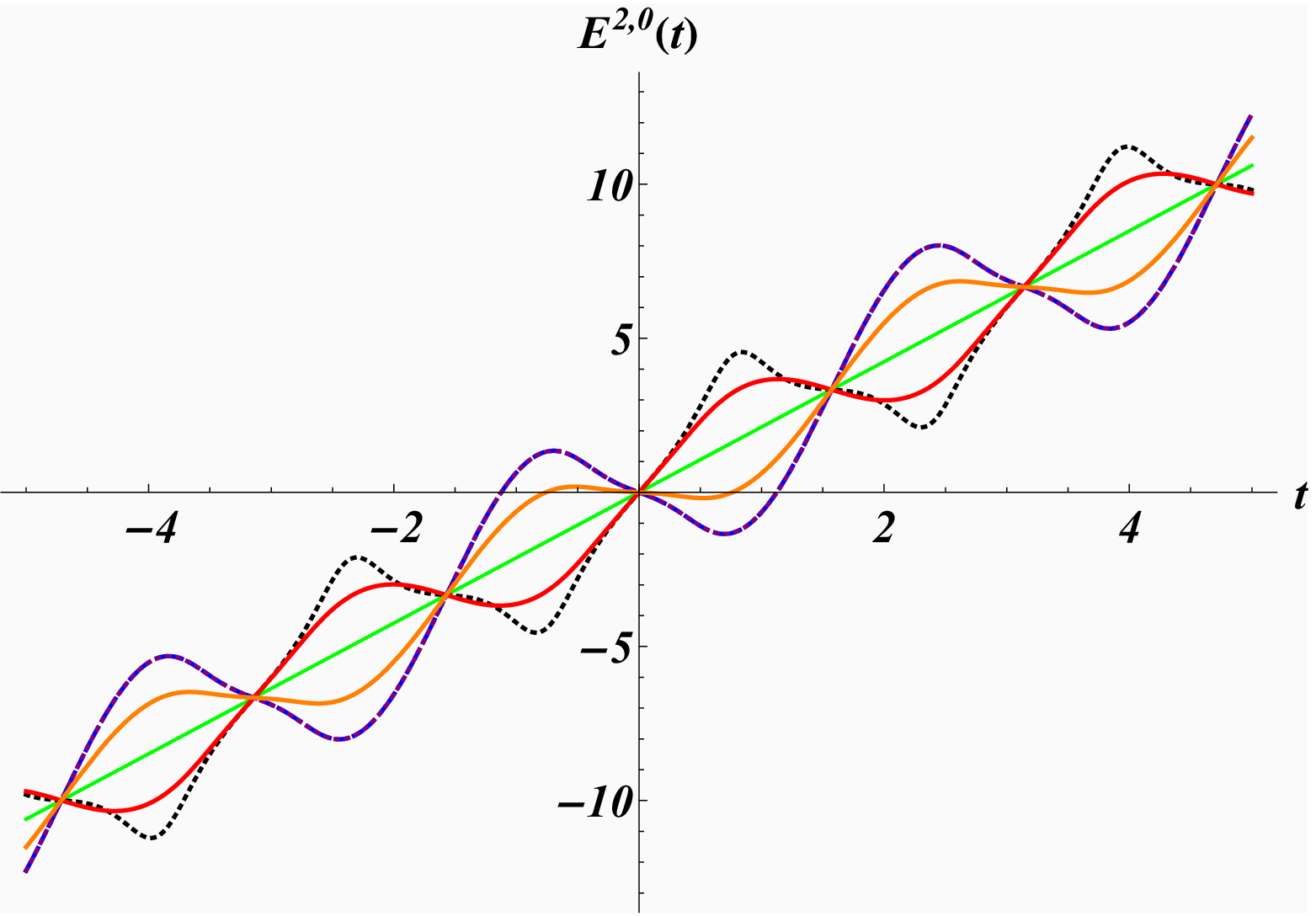, width=6.8cm}
\caption{The instantaneous energy spectra (\ref{inenergy}) associated with the six
Dyson maps for $\lambda (t)=\sin (2t)$ for case 1 with $c_{+}=c_{-}=1$, $k_{1}=2$. In panels (a), (c) we have $a(t)=\cos (t)$ and in panels (a), (c)
we that $a(t)=t/2$.}
       \label{Fig1}}

For concrete choices of the time-dependent coefficient functions we can now
directly evaluate the expressions for $E_{i}^{n,m}(t)$ corresponding to the
Dyson maps $\eta _{i}(t)$ explicitly by computing the auxiliary functions $%
\chi (t)$ and the functions $f_{i}(t)$. The Dyson map $\eta _{2}$ leads to
somewhat different behaviour. This is understood by the fact that it can
only be constructed at $c=0$ and at what would be the exceptional point in
the time-independent scenario $c=\lambda $. Hence also the energies exhibit
slightly different characteristics. Taking the above mentioned constraints
into account there are large regions in the parameter space for which the
all ot the energies $E_{i}^{n,m}(t)$ are real and hence physical. We
illustrate the behaviour of these energies for each of the Dyson maps in
figues \ref{Fig1} and \ref{Fig2} for some concrete choices.

\FIGURE{\epsfig{file=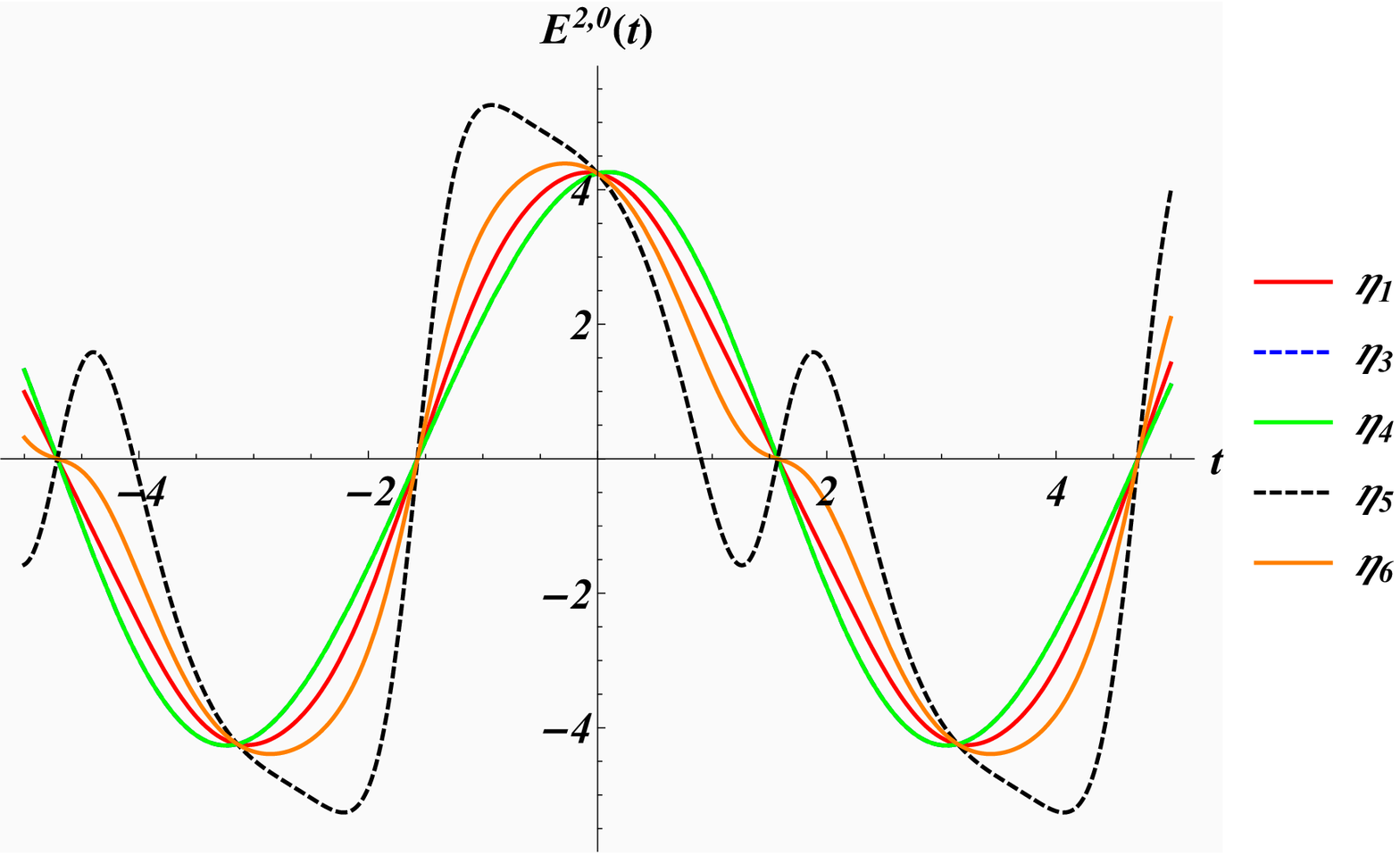, width=7.8cm} \epsfig{file=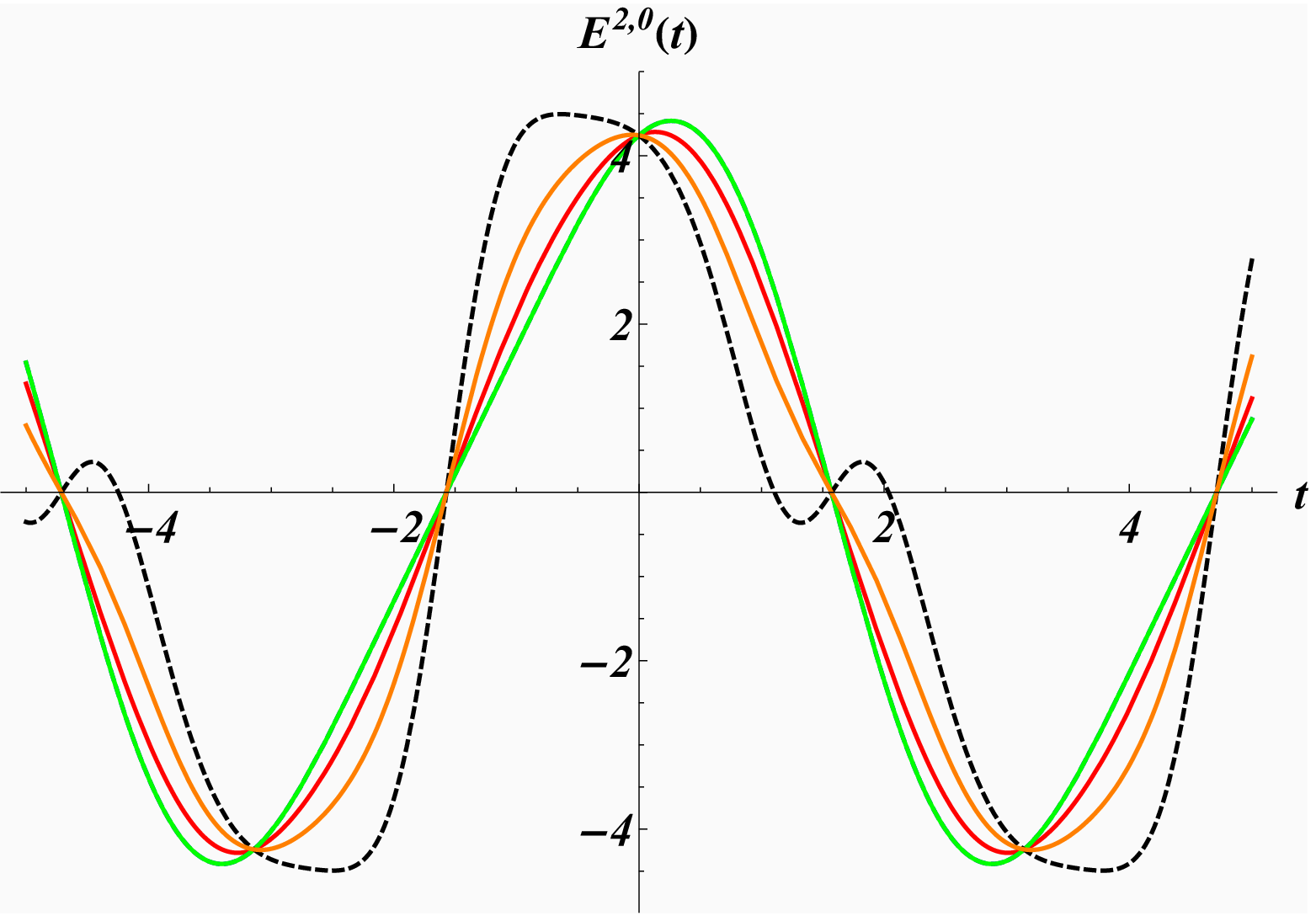, width=6.8cm}
\epsfig{file=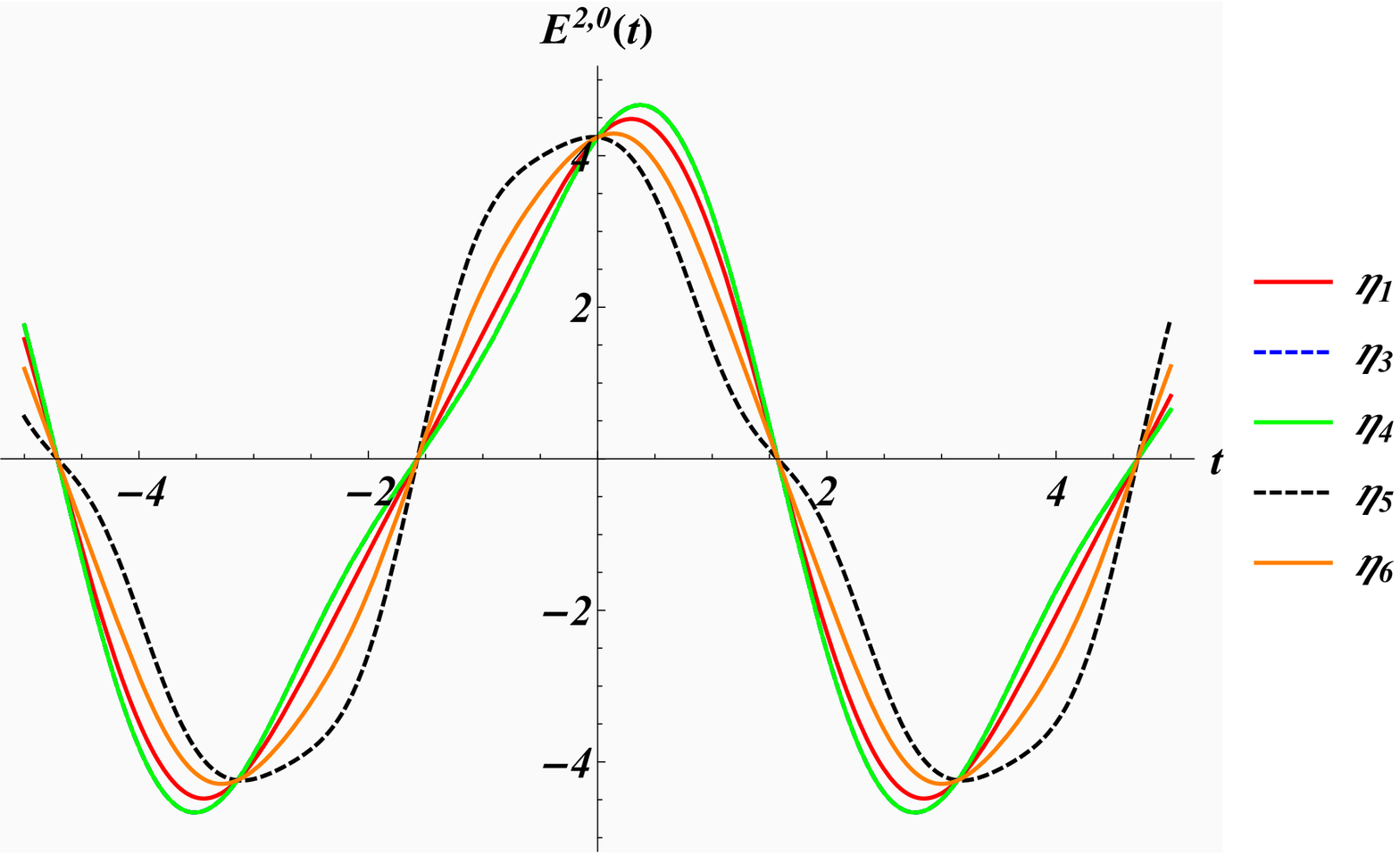, width=7.8cm} \epsfig{file=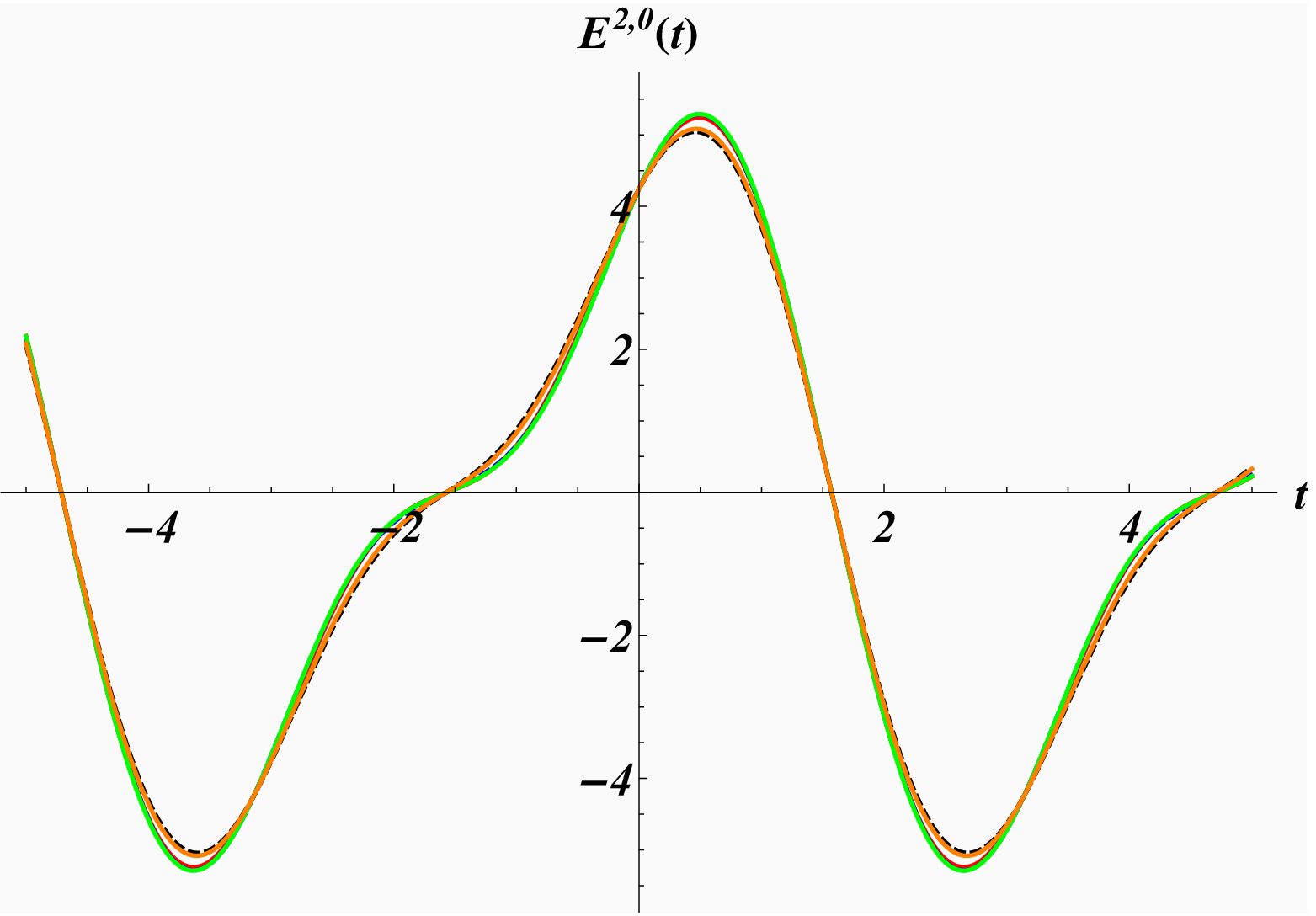, width=6.8cm}
\caption{The instantaneous energy spectra (\ref{inenergy}) associated with five
Dyson maps for $\lambda (t)=\sin (2t)$, $a(t)=\cos(t)$ for case 2 with $c_{+}=c_{-}=1$, $k_{1}=2.5$, $k_{2}=1$. We have $p=-0.1 $, 
$p=-0.3 $, $p=-0.5 $, $p=-0.9 $ in panels (a), (b), (c), (d), respectively.}
       \label{Fig2}}

First of all we observe from figure \ref{Fig1} the crucial feature that the
instantaneous energy is real and finite. Secondly we note that despite
sharing the same non-Hermitian Hamiltonian, the theories related to
different Dyson maps can lead to quite different physical behaviour in the
energy. Similar to the time-independent scenario, this is the known fact
that the Hamiltonian alone does not define a unique definite physical
system, but to define the physics one also needs to specify the metric, i.e.
the Dyson map. We note that some of the energies can become degenerate, $%
E_{1}^{n,n}=E_{2}^{n,n}$, which can however split when $n\neq m$. As is also
expected from the explicit expressions, the differences are more amplified
the larger $\left\vert n-m\right\vert $. In case 2, when we have non
vanishing values of the parameter $p$ these effects are even more amplified
as can be seen in figure \ref{Fig2}. We notice a strong sensitivity with
regard to $p$.

The constraints resulting from the parameterization, $\left\vert
p\right\vert <1$, imply that we are in the regime with spontaneously broken $%
\mathcal{PT}$-symmetry when compared to the time-independent case.
Therefore, we observe the same phenomenon that was first noted in \cite%
{Fring2017,Fring2018b}, namely that the introduction of a time-dependence
into the metric will mend the spontaneously broken $\mathcal{PT}$-regime so
that it becomes physically meaningful. In this case this manifests itself by
the fact that the instantaneous energy is real.

\subsection{Metric and Dyson maps with $\protect\mu (t)\neq 0$, case 3}

Finally we also discuss the case 3 by including a Hermitian coupling term
into the Hamiltonian in addition to the non-Hermitian one. This case turns
out to be more complicated to solve, but may also be tackled successfully by
our perturbative method. Keeping the expression (\ref{anrho}) as our Ansatz
for the perturbative expansion for the metric we obtain the same first order
equation (\ref{eq:1st}), but now involving 
\begin{equation}
h_{0}(t)=a(t)\left( K_{1}+K_{2}\right) +\mu (t)K_{4}\qquad \text{and}\qquad
h_{1}(t)=\lambda (t)K_{3}.
\end{equation}%
Since all generators of the algebra commute with $K_{1}+K_{2}$ the only
nontrivial contribution in the commutator of that relation results from the
term involving $K_{4}$ in $h_{0}$. Taking now 
\begin{equation}
q_{1}=K_{1},\qquad q_{2}=K_{2},\qquad q_{3}=K_{3},  \label{qqq}
\end{equation}%
leads to the following first order equations for the time-dependent
coefficient functions 
\begin{align}
& \dot{\gamma}_{1}^{(1)}(t)=-\frac{1}{2}\mu (t)\gamma _{3}^{(1)}(t), \\
& \dot{\gamma}_{2}^{(1)}(t)=\frac{1}{2}\mu (t)\gamma _{3}^{(1)}(t), \\
& \dot{\gamma}_{3}^{(1)}(t)=\mu (t)\left[ \gamma _{1}^{(1)}(t)-\gamma
_{2}^{(1)}(t)\right] -\lambda (t).  \label{s3}
\end{align}%
We see immediately that $\gamma _{2}^{(1)}(t)=-\gamma _{1}^{(1)}(t)$, which
then also simplifies equations (\ref{s3}).

Proceeding now in the same manner as in the previous cases by extrapolation
to the full series, we find that the following two equations need to be
satisfied 
\begin{equation}
\dot{\gamma}_{1}(t)=-\frac{1}{2}\sinh [\gamma _{3}(t)]\mu (t)\qquad \text{and%
}\qquad \dot{\gamma}_{3}(t)=\cosh [\gamma _{3}(t)]\tanh [2\gamma _{1}(t)]\mu
(t)-\lambda (t).  \label{g13}
\end{equation}%
Letting $\lambda =p\mu $, we can express $\gamma _{3}$ as a function of $%
\gamma _{1}$ 
\begin{equation}
\gamma _{3}(\gamma _{1})=\pm \func{arccosh}\left[ p\tanh (2\gamma _{1})-%
\frac{k_{1}}{2}\func{sech}(2\gamma _{1})\right] .
\end{equation}%
Setting 
\begin{equation}
\gamma _{1}=\frac{1}{2}\func{arcsinh}(\chi ),
\end{equation}%
the two first order equations (\ref{g13}) are converted into the linear
second order auxiliary equation (\ref{EP}) with $\lambda \rightarrow \mu $.
The resulting Hermitian Hamiltonian consists now not only of two decoupled
harmonic oscillators, but also contains an additional Hermitian term in form
of $K_{4}$ 
\begin{equation}
h(t)=a(t)\left( K_{1}+K_{2}\right) -\frac{k_{1}+2p\chi (t)}{2\left[ 1+\chi
(t)^{2}\right] }\mu (t)K_{4}.  \label{h111}
\end{equation}

As in the previous two cases, we may also construct a non-Hermitian solution
for the Dyson map by means of the perturbative approach. From the first
order equation we observe that also $q_{3}=iK_{4}$ with $q_{1}$ and $q_{2}$
as in (\ref{qqq}) leads to a solution. Extrapolating to all orders yields
now the two equations 
\begin{equation}
\dot{\gamma}_{1}(t)=-\frac{1}{2}\sin [\gamma _{3}(t)]\lambda (t)\qquad \text{%
and}\qquad \dot{\gamma}_{3}(t)=\mu (t)-\cos [\gamma _{3}(t)]\coth [2\gamma
_{1}(t)]\lambda (t).  \label{g132}
\end{equation}%
As before we must restrict $\lambda (t)=p\mu (t)$ so that we may solve for $%
\gamma _{3}$ in terms of $\gamma _{1}$ 
\begin{equation}
\gamma _{3}(\gamma _{1})=\pm \arccos \left\{ \frac{\left[ 2-ik_{2}+2\cosh
\left( 2\gamma _{1}\right) \right] \func{csch}(2\gamma _{1})}{2p}\right\} .
\end{equation}%
We set here $k_{2}=0$ in order to obtain a real solution. Letting now%
\begin{equation}
\gamma _{1}=\frac{1}{2}\func{arccosh}(\chi ),
\end{equation}%
the two first order equations (\ref{g132}) are now converted into the linear
second order auxiliary equation (\ref{EP}) with $\lambda \rightarrow \mu $
and $k_{1}\rightarrow 0$. Similarly as the resulting Hamiltonian for the
Hermitian Dyson map the resulting Hermitian Hamiltonian contain a $K_{4}$
besides the two uncoupled harmonic oscillators 
\begin{equation}
h(t)=a(t)\left( K_{1}+K_{2}\right) +\frac{\mu (t)}{\chi (t)-1}K_{4}.
\label{h222}
\end{equation}%
The generator $K_{4}$ can be identified with the standard angular momentum
operator $L_{z}$ and can be eliminated from $h(t)$ in (\ref{h111}) and (\ref%
{h222}) by means of a unitary transformation, see for instance \cite{twoDHO2}%
. Subsequently the eigenfunctions and expectation values of the resulting
system of two uncoupled harmonic oscillators can be obtained similarly as
for the cases $1$ and $2$ presented in detail in the previous section.

\section{The unstable anharmonic quartic oscillator}

In this section we discuss an example for which the previous versions of the
perturbative expressions for the metric or the Dyson map do not however lead
to any solution. In fact, as we will demonstrate one does not only have to
change the Ansatz, but one also needs to rescale the Hamiltonian in order to
introduce the perturbative parameter in the right terms and treat the
non-Hermitian part as a strong rather than a weak perturbation.

Unstable anharmonic oscillators have been the testing ground for
perturbative methods for nonlinear systems for more than fifty years \cite%
{benderwu1969,andrianov1982,graffi1983,caliceti1988,buslaev1993}. Only
fairly recently an exact solution for the time-independent unstable
anharmonic quartic oscillator was found by Jones and Mateo \cite{JM}. They
used ideas from non-Hermitian $\mathcal{PT}$-symmetric quantum mechanics 
\cite{Alirev,PTbook} and applied a perturbative approach that turned out to
be exact. Recently we \cite{BeckyAnd2} also solved the explicitly
time-dependent version of this model in an exact manner. These exact
solutions found in \cite{BeckyAnd2} will serve here as a benchmark for our
perturbative approach, so that we consider the same Hamiltonian, but with
the time-dependent mass term set to zero 
\begin{equation}
H(z,t)=p^{2}-\frac{g(t)}{16}z^{4},~~~~~g\in \mathbb{R}^{+}\text{.}
\label{tanharm}
\end{equation}%
Defining $H(z,t)$ on the contour $z=-2i\sqrt{1+ix}$ as proposed in \cite{JM}%
, it is mapped into the non-Hermitian Hamiltonian 
\begin{equation}
H(x,t)=p^{2}-\frac{1}{2}p+\frac{i}{2}\{x,p^{2}\}+g(t)(x-i)^{2},  \label{H2}
\end{equation}%
where $\{\cdot ,\cdot \}$ denotes as usual the anti-commutator. As mentioned
using our previous versions for the perturbative Ansatz does not lead to a
solvable first order equation or a recursive system. Instead we change our
Ansatz to 
\begin{equation}
\rho (t)=\eta (t)^{\dagger }\eta (t)=\prod_{i=j}^{1}\left[
\prod_{l=k}^{1}\exp \left( \epsilon ^{-l}(\gamma _{i}^{(l)})^{\dagger
}q_{i}\right) \right] \prod_{i=1}^{j}\left[ \prod_{l=1}^{k}\exp \left(
\epsilon ^{-l}(\gamma _{i}^{(l)})q_{i}\right) \right] .  \label{eq:ansatz}
\end{equation}%
As we are expanding in $\epsilon ^{-1}$ we assume here that perturbation
parameter, $\epsilon \gg 1$, is large. The reason for this is that in
addition we also need to scale the Hamiltonian (\ref{H2}) as $x\rightarrow
\epsilon x$. Separating now into a Hermitian and non-Hermitian term, $%
h_{0}(t)$ and $h_{p}(t)$, respectively, we have 
\begin{equation}
h_{0}(t)=p^{2}-\frac{1}{2}p+\epsilon ^{2}g(t)x^{2}-g(t),\qquad \text{and}%
\qquad h_{p}(t)=-2i\epsilon g(t)x+\frac{1}{2}i\epsilon \{x,p^{2}\}.
\end{equation}%
Thus instead of adding a small non-Hermitian perturbation to the Hermitian
part, we have perturbed by a large term and also scaled up the harmonic
oscillator term. Our Hamiltonian acquires therefore the following generic
form 
\begin{equation}
H(t)=h_{1}(t)+\epsilon ^{2}h_{2}(t)+i\epsilon h_{3}(t),
\end{equation}%
which together with the Ansatz (\ref{eq:ansatz}) leads to the new first
order equation 
\begin{equation}
2ih_{3}(t)+\sum_{i=1}^{j}\left[ \left( (\gamma _{i}^{(1)}+(\gamma
_{i}^{(1)})^{\dagger }\right) \left[ q_{i},h_{2}(t)\right] \right] =0.
\label{eq:strongnew}
\end{equation}%
From this equation we can see that if any of the time-dependent coefficient
functions $\gamma _{i}^{(1)}$'s are purely imaginary, then their
contributions vanishes at this order and if they are real we simply acquire
a factor of 2. This version of the Ansatz leads to a recursive system that
can be solved systematically order by order. In our example for the
Hamiltonian (\ref{H2}) we identify 
\begin{equation}
h_{3}(t)=h_{p}(t)\qquad \text{and}\qquad h_{2}(t)=g(t)x^{2},
\end{equation}%
and may satisfy the lowest order equation with the choice 
\begin{equation}
q_{1}=x,\qquad q_{2}=p^{2},\qquad q_{3}=p^{2},\qquad q_{4}=p,
\end{equation}%
where for $q_{3}$ and $q_{4}$ we are taking their time-dependent coefficient
functions to be purely imaginary. In doing so we end up with following
equations that need to be satisfied 
\begin{equation}
\gamma _{2}^{(1)}=\frac{1}{6g},\qquad \text{and}\qquad \gamma _{3}^{(0)}=%
\frac{1}{2\gamma _{1}^{(1)}}.
\end{equation}

At order $\epsilon ^{0}$ we read off the constraining equations 
\begin{equation}
\gamma _{2}^{(2)}=0\qquad \text{and}\qquad \gamma _{1}^{(2)}=-2\left( \gamma
_{1}^{(1)}\right) ^{2}\gamma _{3}^{(1)}.  \label{67}
\end{equation}

\noindent Continuing to order $\epsilon ^{-1}$ we find the constraints%
\begin{equation}
\gamma _{1}^{(1)}=\frac{\dot{g}}{6g},~~\gamma _{1}^{(3)}=-\frac{\gamma
_{3}^{(2)}\dot{g}^{2}}{18g^{2}}+\frac{\dot{g}^{3}}{72g^{4}}+\frac{\left(
\gamma _{3}^{(1)}\right) ^{2}\dot{g}^{3}}{54g^{3}}-\frac{\dot{g}\ddot{g}}{%
72g^{3}},~~\dot{\gamma}_{4}^{(0)}+\gamma _{4}^{(0)}\left( \frac{\ddot{g}}{%
\dot{g}}-\frac{\dot{g}}{g}\right) =-\frac{1}{3}.  \label{drei}
\end{equation}%
The last equation is solved to 
\begin{equation}
\gamma _{4}^{(0)}=\frac{c_{1}g}{\dot{g}}-\frac{g\log {g}}{2\dot{g}}.
\end{equation}%
At order $\epsilon ^{-2}$ we obtain $\gamma _{3}^{(1)}=0$, and therefore
with (\ref{67}) we have $\gamma _{1}^{(2)}=0$.

\noindent At order $\epsilon ^{-3}$ we obtain 
\begin{equation}
\gamma _{3}^{(2)}=\frac{\dot{g}^{2}-g\ddot{g}}{4g^{2}\dot{g}},
\end{equation}%
which implies with (\ref{drei}) that $\gamma _{1}^{(3)}=0$. Some features
hold for all remaining orders in $\varepsilon $. We have $\gamma
_{2}^{(i)}=0 $ for all $i\geq 2$. We also find that at every order $\epsilon
^{-n}$, where $n\geq 2$ the differential equation 
\begin{equation}
\frac{\gamma _{4}^{(n-1)}\dot{g}^{2}}{3g^{2}}+\frac{\dot{g}\dot{\gamma}%
_{4}^{(n-1)}}{3g}+\frac{\gamma _{4}^{(n-1)}\ddot{g}}{3g}=0,
\end{equation}%
occurs, which is solved by 
\begin{equation}
\gamma _{4}^{(n-1)}=\frac{c_{n-1}g}{\dot{g}}
\end{equation}%
Another equation that appears at all orders $\epsilon ^{-n}$ for $n\geq 2$
is given by 
\begin{equation}
\gamma _{1}^{(n+2)}=-\frac{\gamma _{3}^{(n+1)}\dot{g}^{2}}{18g^{2}}
\end{equation}%
This is solved at all orders if we have 
\begin{equation}
\gamma _{1}^{(n+2)}=0\qquad \text{and}\qquad \gamma _{3}^{(n+1)}=0
\end{equation}%
for $n\geq 2$. When eliminating the $\gamma $s from these equations we are
left with a differential equation entirely in $g$ given by 
\begin{equation}
\frac{14\dot{g}^{3}}{9g^{2}}+\frac{2\dot{g}\ddot{g}}{g}-\frac{\dddot{g}}{2}=0
\end{equation}%
Parameterizing $g=\frac{1}{2}\sigma ^{-3}$ this equation reduces to 
\begin{equation}
\sigma ^{2}\dddot{\sigma}=0
\end{equation}%
which is easily solved by $\sigma (t)=c_{1}+c_{2}t+c_{3}t^{2}$.

Assembling all our results we extrapote to all orders, i.e. an exact
solution. Setting therefore $\varepsilon =1$ gives the time-dependent Dyson
map of the form 
\begin{equation}
\eta (t)=\exp [\gamma _{1}(t)x]\exp [\gamma _{2}(t)p^{3}+i\gamma
_{3}(t)p^{2}+i\gamma _{4}(t)p],
\end{equation}%
with 
\begin{equation}
\gamma _{1}=\frac{\dot{g}}{6g},\quad \gamma _{2}=\frac{1}{6g},\quad \gamma
_{3}=\frac{12g^{3}+\dot{g}^{2}-g\ddot{g}}{4\dot{g}g^{2}},\quad \gamma _{4}=%
\frac{g}{\dot{g}}\left( c_{1}-\frac{\log {g}}{2}\right) ,
\end{equation}%
which is in precise agreement with the Dyson map we previously found in \cite%
{BeckyAnd2}.

\section{Conclusions}

We have demonstrated how to set up a perturbative approach that allows to
construct the metric operator and the Dyson map in a recursive manner order
by order in a perturbative parameter that may be very small or very large.
We found three different types of perturbative expansions. The Ansatz (\ref%
{an}) is the most natural one when the Dyson map is assumed to be Hermitian
and needs to be slightly modified when one allows $\eta $ to be
non-Hermitian as shown in section 3.1.2. In both of these versions the
non-Hermitian term was treated as a small perturbation. In section 4 we
demonstrated that this approach can not be applied universally and has to be
altered for some models for which one needs to treat the non-Hermitian term
and parts of the Hermitian term as large perturbations. Consequently the
perturbative expansion needs to be in the inverse of the large perturbative
parameter.

When compared to the time-independent scenario, all our approaches have in
common that the order-by-order equations do not just determine the
commutative structure of the $q_{i}$s, but computations are more involved as
in addition one needs to solve coupled sets of differential equations for
the time-dependent coefficient functions. Moreover, we observed that the key
structure is already determined by the lowest order equation.

Although the main emphasis in this paper is on the perturbation theory, with
regard to the specific example studied we found many new Dyson maps for the
coupled non-Hermitian harmonic oscillator. We saw that these different maps
lead to different types of physical behaviour, as shown explicitly for the
time-dependent energy expectation values. When compared to the
time-independent case, all our solutions are only valid in what would be the
spontaneously broken $\mathcal{PT}$-regime, except for one example that is
defined on what would be the exceptional point. So similar to the effect
observed in \cite{Fring2017, Fring2018b}, this regime becomes physically
meaningful in the time-dependent setting. However, unlike as in some of the
previously studied systems, one can not crossover to the $\mathcal{PT}$%
-regime and is confined to the broken phase. It remains an open issue to
formulate general criteria that characterize precisely when this possibility
occurs for time-dependent systems and when not.

\medskip

\noindent \textbf{Acknowledgments:} RT is supported by a City, University of
London Research Fellowship.

\end{document}